\documentclass[aps,twocolumn,showpacs,preprintnumbers,floatfix,nofootinbib]{revtex4-1}
\usepackage{graphicx}% Include figure files
\usepackage[subfigure]{graphfig}
\usepackage{epsfig}
\usepackage{epstopdf}
\usepackage{dcolumn}% Align table columns on decimal point
\usepackage{amsmath}

\usepackage{xcolor}
\definecolor{darkred}{rgb}{0.4,0.0,0.0}
\definecolor{darkgreen}{rgb}{0.0,0.4,0.0}
\definecolor{darkblue}{rgb}{0.0,0.0,0.4}
\usepackage[bookmarks,linktocpage,colorlinks,
    linkcolor = darkred,
    urlcolor  = darkblue,
    citecolor = darkgreen]{hyperref}
\usepackage{subfigure}
\def\be{\begin{equation}}
\def\ee{\end{equation}}
\def\bea{\begin{eqnarray}}
\def\eea{\end{eqnarray}}
\newcommand{\Tr}{\mathrm {Tr}}

\newcommand{\pslash}[1]{#1\!\!\!/}

\definecolor{green}{rgb}{0,.5,0}

\begin{document}
%%%%%%%%%%%%%%%%%%%%%%%%%%%%%%%%%%%%%%%%%%%%%%%%%%%%%%%%%%%%%%%%%%%%%%%%%%%%%
%
%
\title{\vspace{1.0in} {\bf Proton Mass Decomposition from the QCD Energy Momentum Tensor}} 

\author{Yi-Bo Yang$^{1}$,  Jian Liang$^{2}$, Yu-Jiang Bi$^{3}$, Ying Chen$^{3,4}$, Terrence Draper$^{2}$, Keh-Fei Liu$^{2}$ and Zhaofeng Liu$^{3,4}$
\vspace*{-0.5cm}
\begin{center}
\large{
\vspace*{0.4cm}
\includegraphics[scale=0.20]{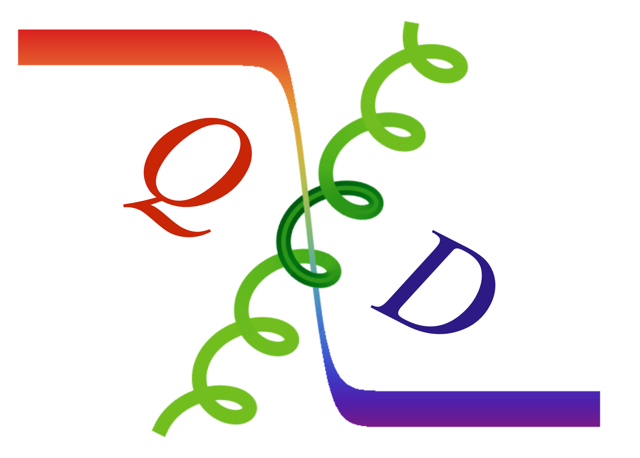}\\
\vspace*{0.4cm}
($\chi$QCD Collaboration)
}
\end{center}
}
\affiliation{
$^{1}$\mbox{Department of Physics and Astronomy, Michigan State University, East Lansing, MI 48824, USA}\\
$^{2}$\mbox{Department of Physics and Astronomy, University of Kentucky, Lexington, KY 40506, USA}\\
$^{3}$\mbox{Institute of High Energy Physics, Chinese Academy of Sciences, Beijing 100049, China}\\
$^{4}$\mbox{School of Physics, University of Chinese Academy of Sciences, Beijing 100049, China}\\
}

\preprint{MSUHEP-18-017}

%----------------------------------------------------------------------------
\begin{abstract}
We report results on the proton mass decomposition and also on related quark and glue momentum fractions. The results are based on overlap valence fermions on four ensembles of $N_f = 2+1$ DWF configurations with three lattice spacings and three volumes, and several pion masses including the physical pion mass. With fully non-perturbative renormalization (and universal normalization on both quark and gluon), we find that the quark energy and glue field energy contribute 33(4)(4)\% and 37(5)(4)\% respectively in the $\overline{MS}$ scheme at $\mu = 2$ GeV.  A quarter of the  trace anomaly gives a 23(1)(1)\% contribution to the proton mass based on the sum rule, given 9(2)(1)\% contribution from the $u, d,$ and $s$ quark scalar condensates. The $u,d,s$ and glue momentum fractions in the $\overline{MS}$ scheme are in good agreement with global analyses at $\mu = 2$ GeV.
\end{abstract}
\maketitle

{\it Introduction: }
In the standard model, Higgs boson provides the origin of quark masses. But how it is related to the proton mass and thus the masses of nuclei and atoms is another question. The masses of the valence quarks in the proton are just $\sim$3 MeV per quark which is directly related to the Higgs boson, while the total proton mass is 938 MeV. The percentage of the quark and gluon contributions to the proton mass can only be provided by solving QCD non-perturbatively, and/or with information from experiment. With phenomenological input, the first decomposition was carried out by Ji~\cite {Ji:1994av}. As in Refs.~\cite{Ji:1994av,Yang:2014xsa}, the Hamiltonian of QCD can be decomposed as
 \be 
 M = - \langle T_{44} \rangle =\langle H_m\rangle + \langle H_E\rangle (\mu)+ \langle H_g\rangle (\mu) +  \frac{1}{4} \langle H_a \rangle, \label{eq:total}
\ee
in the rest frame of the hadron state where M is the hadron mass, $T_{\mu\nu}$ is the energy momentum tensor of QCD with $ \langle T_{44} \rangle$ as its expectation value in the hadron, and the trace anomaly gives
\be
M = - \,\langle \hat{T}_{\mu\mu} \rangle= \langle H_m\rangle +  \,\langle H_a\rangle.\label{eq:anomaly}
\ee
The $H_m$, $H_E$, and $H_g$ in the above equations denote the contributions from the quark condensate, the quark energy, and the glue field energy, respectively:
\bea
  H_m &=& \sum_{u,d,s\cdots}\int d^3x\, m\, \overline \psi  \psi,  
  H_E = \sum_{u,d,s...}\int d^3x~\overline \psi(\vec{D}\cdot \vec{\gamma})\psi,\quad  \nonumber\\
%H_q = H_E+H_m&, \quad
 H_g &=& \int d^3 x~ {\frac{1}{2}}(B^2- E^2).
 \eea
The QCD anomaly term $H_a$ is the joint contribution from the quantum anomaly of both glue and quark,
 \bea
 H_a= H_g^a +H^{\gamma}_m, \quad
 H_g^a &=&\int d^3x~ \frac{-\beta(g)}{g}( E^2+ B^2), \nonumber\\\quad
H^{\gamma}_m&=&\sum_{u,d,s\cdots}\int d^3x\, \gamma_m m\, \overline \psi  \psi.
 \eea
All the $\langle H \rangle$ are defined by $\langle N|H |N\rangle/\langle N|N\rangle$ where $|N\rangle$ is the nucleon state in the rest frame.
Note that $\langle H_E+H_g\rangle$, $\langle H_m\rangle$ and $\langle H_a\rangle$ are scale and renormalization scheme independent, but 
$\langle H_E\rangle (\mu)$ and $\langle H_g\rangle (\mu)$ separately have scale and scheme dependence.

The nucleon mass $M$ can be calculated from the nucleon two-point function. If one calculates
further $\langle H_m\rangle$ and $\langle H_E\rangle (\mu)$, then $\langle H_g\rangle (\mu)$ and $\langle H_a\rangle$ can be obtained through Eqs.~(\ref{eq:total}) and (\ref{eq:anomaly}). The approach has been adopted to decompose the S-wave meson masses to gain insight about
contributions of each term from light mesons to charmoninums~\cite{Yang:2014xsa}. But the mixing between $\langle H_E\rangle (\mu)$ and $\langle H_m\rangle$ will be non-trivial under the lattice regularization, when there is any breaking of the quark equation of motion at finite spacing. On the other hand, if we obtain the renormalized quark momentum fraction $\langle x \rangle^R_q$ in the continuum limit, and define the renormalized quark energy $\langle H_E^R\rangle$ in term of $\langle x \rangle^R_q$ and $\langle H_m\rangle$ with the help of the equation of motion, i.e.,
\be    \label{eq:H_E}
\langle H_E^R \rangle =\frac{3}{4} \langle x \rangle^R_q M-\frac{3}{4}\langle H_m\rangle,
\ee
then the additional mixing can be avoided. 
Similarly, the renormalized glue field energy can be accessed from the glue momentum fraction $\langle x \rangle^R_g$ by
\be  \label{eq:H_g}
\langle H^R_g \rangle =\frac{3}{4} \langle x \rangle^R_g M.
\ee

In the present work, we use the lattice derivative operator for the quark EMT and combination of plaquettes for the gauge
EMT and address their normalization in addition to renormalization and mixing. We calculate the proton mass  and the renormalized $\langle x \rangle_{q,g}$ on four lattice ensembles, and extrapolate the results to the physical pion mass with a global fit including finite lattice spacing and volume corrections. Then we combine previously calculated $\langle H_m\rangle$~\cite{Yang:2015uis} to obtain $\langle H_a\rangle$ from Eq.~(\ref{eq:anomaly}), and the full decomposition of the proton energy in the rest frame as shown in Eq.~(\ref{eq:total}). 
\\

{\it Numerical setup: } We use overlap valence fermions on $(2+1)$ flavor RBC/UKQCD DWF gauge configurations from four ensembles on $24^3\times64$ (24I),  $32^3\times64$ (32I)~\cite{Aoki:2010dy}, $32^3\times64$ (32ID) and $48^3\times96$ (48I)~\cite{Blum:2014tka} lattices.  These ensembles cover three values of the lattice spacing and volume respectively, and four values of the quark mass in the sea, which allows us to implement a global fit on our results to control the systematic uncertainties as in Ref.~\cite{Yang:2015uis,Sufian:2016pex}. Other parameters of the ensembles used are listed in Table~\ref{table:r0}. 

\begin{table}[htbp]
\begin{center}
\caption{\label{table:r0} The parameters for the RBC/UKQCD configurations~\cite{Blum:2014tka}: spatial/temporal size, lattice spacing, sea strange quark mass under $\overline{\textrm{MS}}$ scheme at {2 GeV}, pion mass with the degenerate light sea quark, and the number of configurations.}
\begin{tabular}{ccccccc}
Symbol & $L^3\times T$  &a (fm)  &$m_s^{(s)}$(MeV)&  {$m_{\pi}$}(MeV)   & $N_{cfg} $ \\
\hline
32ID &$32^3\times 64$& 0.1431(7) & 89.4   &171 & 200  \\
24I & $24^3\times 64$& 0.1105(3) &120   &330  & 203  \\
48I &$48^3\times 96$& 0.1141(2) & 94.9   &139 & 81  \\
32I &$32^3\times 64$& 0.0828(3) & 110   &300 & 309  \\
\hline
\end{tabular}
\end{center}
\end{table}

The effective quark propagator of the massive
overlap fermion is the inverse of the operator $(D_c + m)$~\cite{Chiu:1998eu,Liu:2002qu}, where  $D_c$ is chiral, i.e. $\{D_c, \gamma_5\} = 0$ \cite{Chiu:1998gp} and its detailed definition can be found in our previous work~\cite{Li:2010pw,Gong:2013vja,Yang:2015zja}. We used 4 quark {masses} from the range $m_{\pi}\in$(250,400) MeV on the 24I and 32I ensembles, and 6/5 quark 
masses from $m_{\pi}\in$(140,400) MeV on the 48I/32ID ensemble respectively which have larger volumes and thus allow a lighter pion mass with 
the constraint $m_{\pi}L>3.8$.  One step of HYP smearing is applied on all the configurations to improve the signal. Numerical details regarding the calculation of the overlap operator, eigenmode deflation in inversion of the quark matrix, and the $Z_3$ grid smeared source with low-mode substitution (LMS) to increase statistics are given in~\cite{Li:2010pw,Gong:2013vja,Yang:2015zja,Liang:2016fgy}.

 \begin{figure}[h]
\begin{center}
    \includegraphics[scale=0.5]{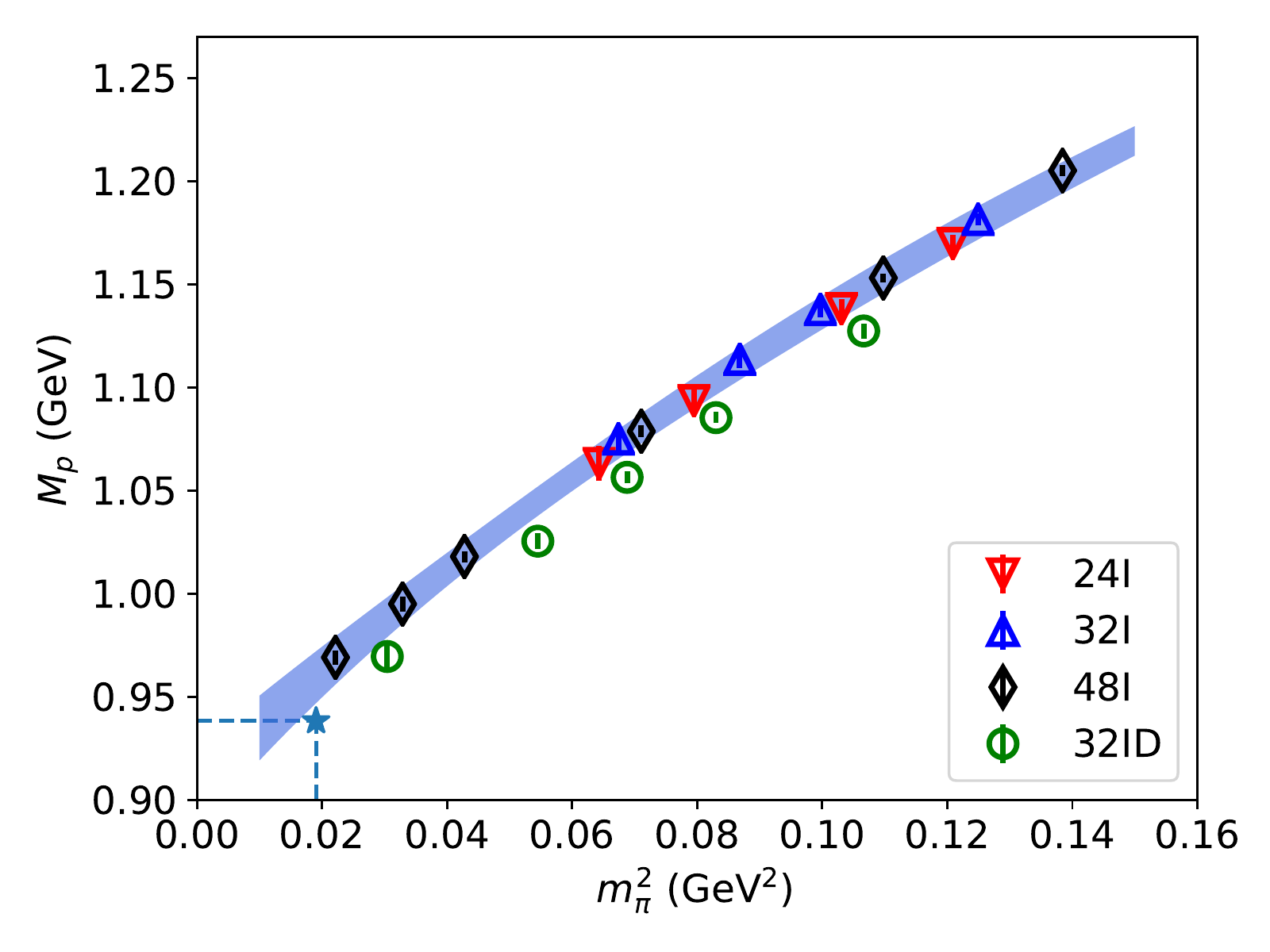}
\end{center}
\caption{The proton mass as a function of the pion mass at different lattice spacings and volumes, after partially quenching effects are subtracted. The star shows the physical proton mass.}
\label{fig:mass}
\end{figure}

{\it Proton mass:} We first calculate the proton mass on these four ensembles and apply the SU(4$\vert$2) mixed action HB$\chi$PT functional form~\cite{Tiburzi:2005is} to fit the results,
\bea
M(m_{\pi}^v,m_{\pi}^{sea},a,L)&=&M_0+C_1(m_{\pi}^v)^2+C_2(m_{\pi}^{sea})^2\nonumber\\
&-&\frac{(g^2_A-4g_Ag_1-5g^2_1)\pi}{3(4\pi f_{\pi})^2}(m_{\pi}^{v})^3\nonumber\\
&-&\frac{(8g^2_A+4g_Ag_1+5g^2_1)\pi}{3(4\pi f_{\pi})^2}(m_{\pi}^{pq})^3\nonumber\\
&+&C^{I/ID}_{3} a^2+C_4\frac{(m_{\pi}^v)^2}{L} e^{-m_{\pi}^vL},
\eea
where $M_0$, $C_{1,2,3,4}$, the axial vector coupling $g_A$ and an additional partially quenched one $g_1$ are free parameters, $f_{\pi}=0.122(9)$~GeV is the pion decay constant, $m_{\pi}^{v,sea}$ is the valence/sea pion mass respectively, $m_{\pi}^{pq}=\sqrt{(m_{\pi}^{v})^2+(m_{\pi}^{sea})^2+\Delta_{mix}a^2}$ is the partially quenched mass with the mixed action term $\Delta_{mix}a^2$, and $a$ is the lattice spacing. The ${\cal O}(m_{\pi}^3)$ logarithm function ${\cal F}$ in the original  functional form is dropped since it turns out to be not useful to constrain the fit. Note that we used $C^{I}_{3}$ for the 24I/48I/32I ensembles and $C^{ID}_{3}$ for 32ID ensemble as they used different gauge actions. We get the prediction of the proton mass at the physical point as $M(m_{\pi}^{phys},m_{\pi}^{phys},0,\infty)$=0.960(13)~GeV with $\chi^2$/d.o.f.=0.52. From the fit, we can also get the light quark mass sigma term $H_{m,u+d}\simeq \frac{\partial M}{\partial m_{\pi}}m_{\pi}/2=52(8)$~MeV which is consistent with our previous direct calculation 46(7)(2)~MeV~\cite{Yang:2015uis}. The $g_A$ we get from the fit is 0.9(2) which is consistent with the experimental result 1.2723(23)~\cite{Patrignani:2016xqp} within 2$\sigma$. Alternatively, using the experimental value of $g_A$ predicts the proton mass as 0.931(8) with a $\chi^2$/d.o.f. of 1.5. The results of the proton mass with the partially quenching effect ($m_{\pi}^{sea}\neq m_{\pi}^{v}$) subtracted are plotted in Fig.~\ref{fig:mass} as a function of the valence pion mass, together with the blue band for our prediction in the continuum limit. The difference between the results with different symbols reflects the discretization errors and finite volume effects which are reasonably small, as shown in Fig.~\ref{fig:mass}.

{\it Momentum fraction:} The quark and gluon momentum fractions in the nucleon can be defined by the traceless diagonal part of the EMT matrix element in the rest frame~\cite{Horsley:2012pz},
\bea
\langle x \rangle_{q,g}&\equiv&-\frac{\langle N |\frac{4}{3}\overline{T}^{q,g}_{44} |N \rangle}{M\langle N|N \rangle},\\
\overline{T}^{q}_{44}&=&\int d^3x \overline{\psi}(x) \frac{1}{2}(\gamma_4\overleftrightarrow{D}_4 -\frac{1}{4}{\displaystyle\sum_{i=0,1,2,3}} \gamma_i\overleftrightarrow{D}_i) \psi (x), \nonumber\\ \ \ 
\overline{T}^{g}_{44}&=&\int d^3x \frac{1}{2}(E(x)^2-B(x)^2).\nonumber
\eea
 In practice, we calculated ratios of the three-point function to the two-point function
\begin{eqnarray}   \label{eq:ratio}
R^{q,g}(t_f,t)&=&\frac{\langle 0|\int d^3 y\Gamma^e{\chi}_S(\vec{y},t_f)\overline{T}^{q,g}_{44}(t)\sum_{\vec{x}\in G}\bar{\chi}_S(\vec{x},0)|0 \rangle}{\langle 0|\int d^3 y\Gamma^e\chi_S(\vec{y},t_f)\sum_{\vec{x}\in G}\bar{\chi}_S(\vec{x},0)|0 \rangle},\nonumber\\
\end{eqnarray}
where $\chi_S$ is the standard proton interpolation field with Gaussian smearing applied to all three quarks, and $\Gamma^e$ is the unpolarized projection operator of the nucleon. All the correlation functions from the source points $\vec{x}$ in the grid $G$ are combined to improve the the signal-to-noise ratio (SNR)~\cite{Yang:2015zja}. When $t_f$ is large enough, 
 $R^{q,g}(t_f,t)$ approaches the bare nucleon matrix element $\langle N| \overline{T}^{q,g}_{44}|N\rangle$.

For each quark mass on each ensemble, we construct $R(t_f,t)$
 for several sink-source separations $t_f$ from
0.7 fm to 1.5 fm and all the current insertion times
$t$ between the source and sink, combine all the data to do the two-state fit, and then obtain the matrix elements we want with the excited-states contamination removed properly. The more detailed discussion of the simulation setup and the two-state fit can be found in our previous work~\cite{Yang:2015uis,Sufian:2016pex,Yang:2016plb}.

To improve the signal in the disconnected insertion part of $\langle x \rangle_{q,g}$, all the time slices are looped over for the proton propagator. For $\langle x \rangle_{g}$, the cluster-decomposition error reduction (CDER) technique is applied, as described in Refs.~\cite{Liu:2017man,Yang:2018bft}.\\

\begin{widetext}
The renormalized momentum fractions $\langle x \rangle^R$ in the $\overline{\textrm{MS}}$ scheme at scale $\mu$ are
\bea
\langle x \rangle^R_{u,d,s}=Z^{\overline{\textrm{MS}}}_{QQ}(\mu)\langle x \rangle_{u,d,s}+\delta Z^{\overline{\textrm{MS}}}_{QQ}(\mu)\sum_{q=u,d,s} \langle x \rangle_{q}+Z^{\overline{\textrm{MS}}}_{QG}(\mu)\langle x \rangle_g,\ \langle x \rangle^R_g=Z^{\overline{\textrm{MS}}}_{GQ}(\mu)\sum_{q=u,d,s} \langle x \rangle_{q}+Z^{\overline{\textrm{MS}}}_{GG}\langle x \rangle_g,
\eea
where $\langle x \rangle_{u,d,s,g}$ is the bare momentum fraction under the lattice regularization, and 
the renormalization constants in the $\overline{\textrm{MS}}$ at scale $\mu$ are defined through the RI/MOM scheme 
\bea
&&\left(\begin{array}{cc}
Z^{\overline{\textrm{MS}}}_{QQ}(\mu)+N_f\delta Z^{\overline{\textrm{MS}}}_{QQ} (\mu)&
N_fZ^{\overline{\textrm{MS}}}_{QG}(\mu)   \\
Z^{\overline{\textrm{MS}}}_{GQ}(\mu) &
Z^{\overline{\textrm{MS}}}_{GG}(\mu)
\end{array}\right)\equiv\left\{\left[\left(\begin{array}{cc}
Z_{QQ}(\mu_R)+N_f\delta Z_{QQ}&
N_fZ_{QG}(\mu_R)\\
Z_{GQ}(\mu_R)&
Z_{GG}(\mu_R)
\end{array}\right)\right.\right.\nonumber\\
&&\quad\quad\quad\quad\quad\quad\quad\quad\quad\quad\quad\quad\quad\quad \left.\left.
\left(\begin{array}{cc}
R_{QQ}(\frac{\mu}{\mu_{R}})+{\cal O}(N_f\alpha_s^2) &
N_fR_{QG}(\frac{\mu}{\mu_{R}})\\
R_{GQ}(\frac{\mu}{\mu_{R}}) &
R_{GG}(\frac{\mu}{\mu_{R}})
\end{array}\right)
\right]|_{a^2\mu_R^2\rightarrow0}\right\}^{-1}
\eea
and $Z_{QQ}(\mu)=\left[\left(Z_{QQ}(\mu_R)R_{QQ}(\mu/\mu_R)\right)|_{a^2\mu_R^2\rightarrow 0}\right]^{-1}$. Note that the iso-vector matching coefficient $R_{QQ}(\frac{\mu}{\mu_{R}})$ has been obtained at the 3-loop level~\cite{Gracey:2003mr} while just the 1-loop level results of the other $R$'s are available~\cite{Yang:2016xsb}. 
\end{widetext}

We list the renormalization constants for $\overline{T}^{q,g}_{44}$ at $\overline{\textrm{MS}}$ \mbox{2~GeV} in Table.~\ref{table:Z} and the details of the NPR calculation are discussed in the supplementary materials~\cite{yang:2018NPR}.

\begin{table}[htbp]
\begin{center}
\caption{\label{table:Z} The non-perturbative renormalization constants on different ensembles,  at $\overline{\textrm{MS}}$ \mbox{2~GeV}. The 24I and 48I ensembles have the same lattice spacing and thus share the renormalization constants. The wwo uncertainties are the statistical and systematic ones respectively with the details provided in the supplementary materials~\cite{yang:2018NPR}.}
\begin{tabular}{cccccc}
Symbol & $Z_{QQ}$  & $\delta Z_{QQ}$ & $Z_{QG}$ & $Z_{GQ}$ &$Z_{GG}$ \\
\hline
32ID & 1.25(0)(2) & 0.018(2)(2) & 0.017(17) &  0.57(3)(6) & 1.29(5)(9) \\
24I/48I & 1.24(0)(2) & 0.012(2)(2) & 0.007(14) & 0.35(3)(6) & 1.07(4)(4) \\
32I &  1.25(0)(2) & 0.008(2)(2) & 0.000(14) & 0.18(2)(2) & 1.10(4)(5) \\
\hline
\end{tabular}
\end{center}
\end{table}

\begin{figure*}[htb!] 
\includegraphics[scale=0.5]{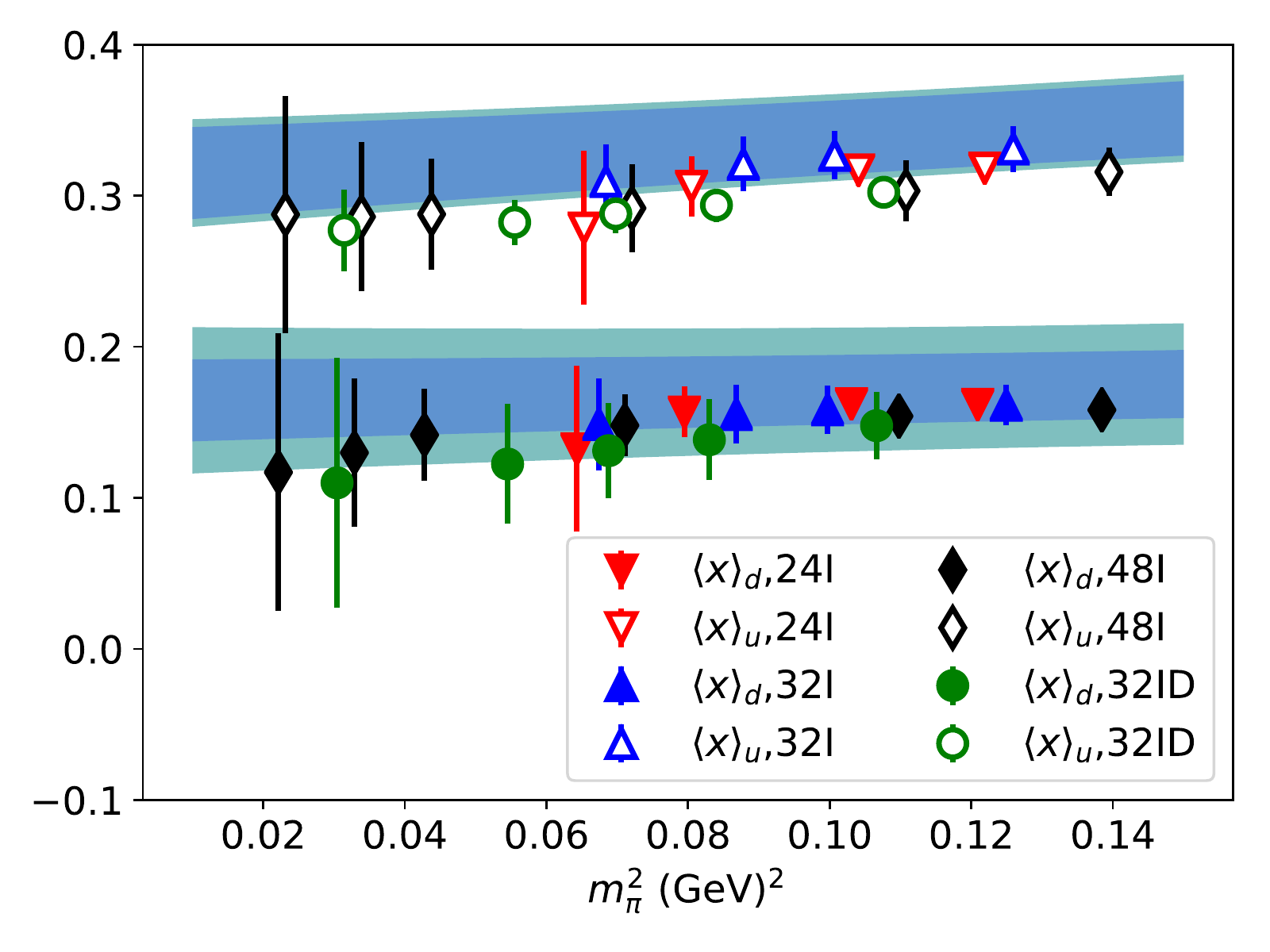}
\includegraphics[scale=0.5]{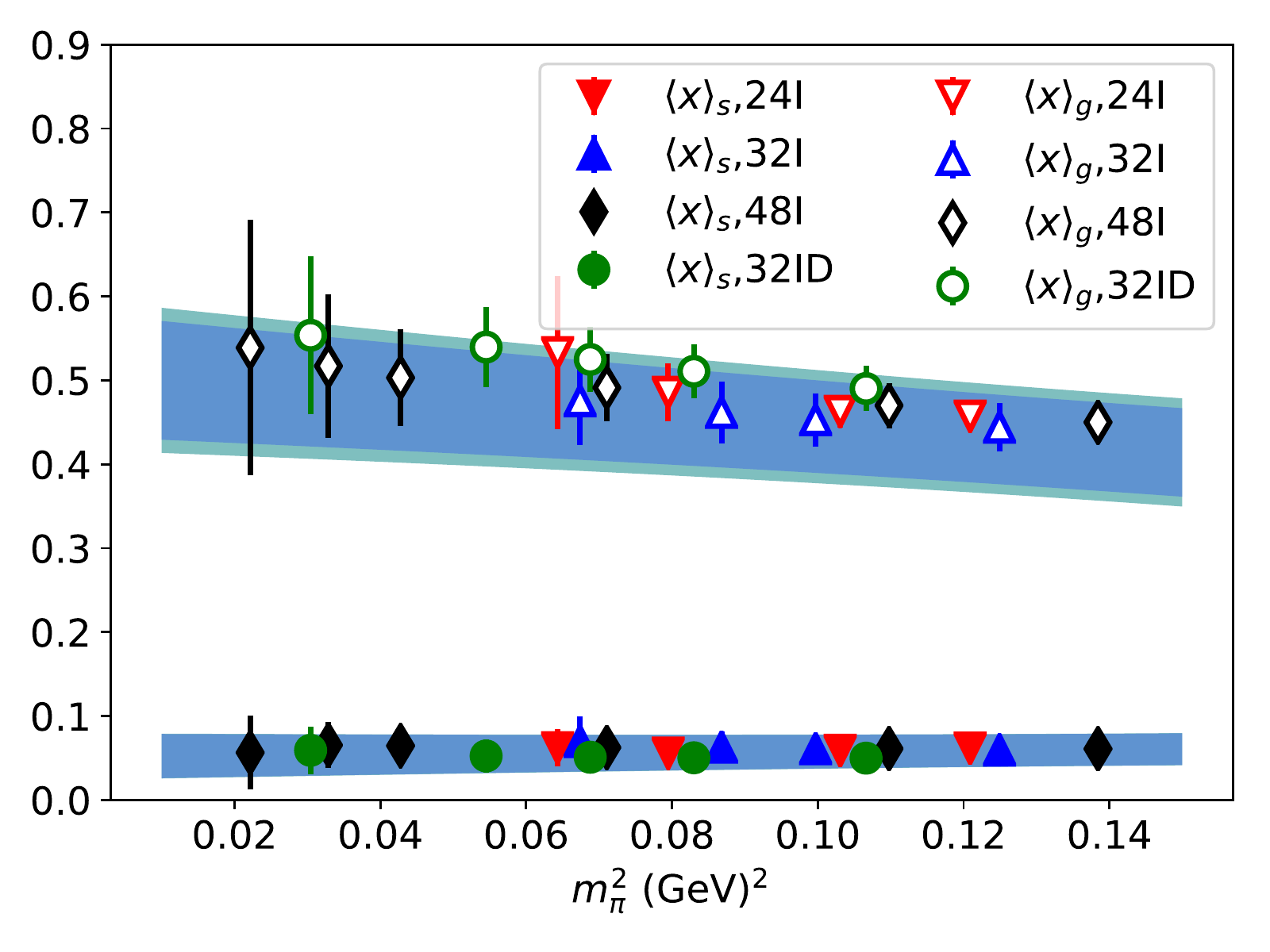}\\
\includegraphics[scale=0.5]{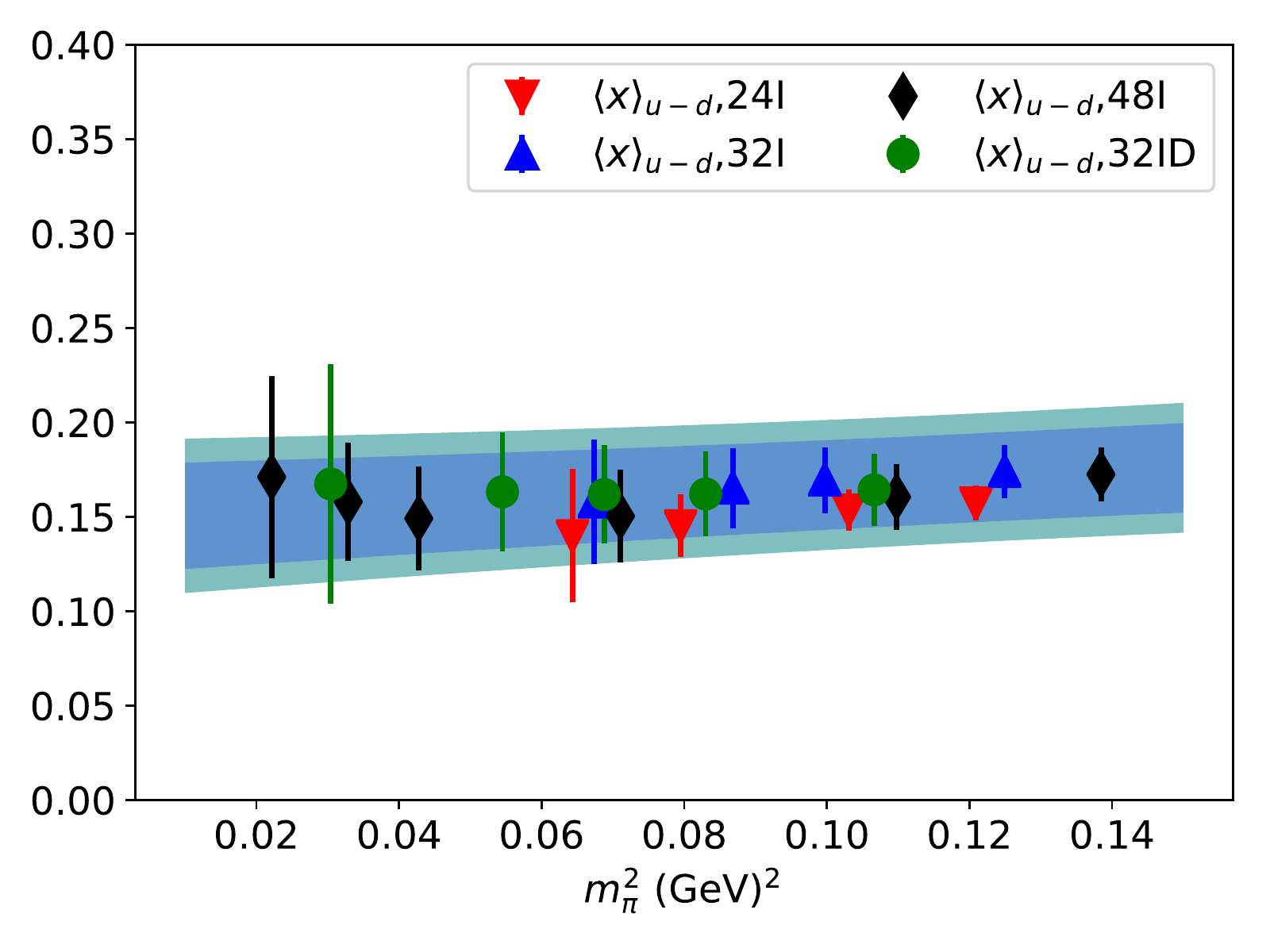}
\includegraphics[scale=0.5]{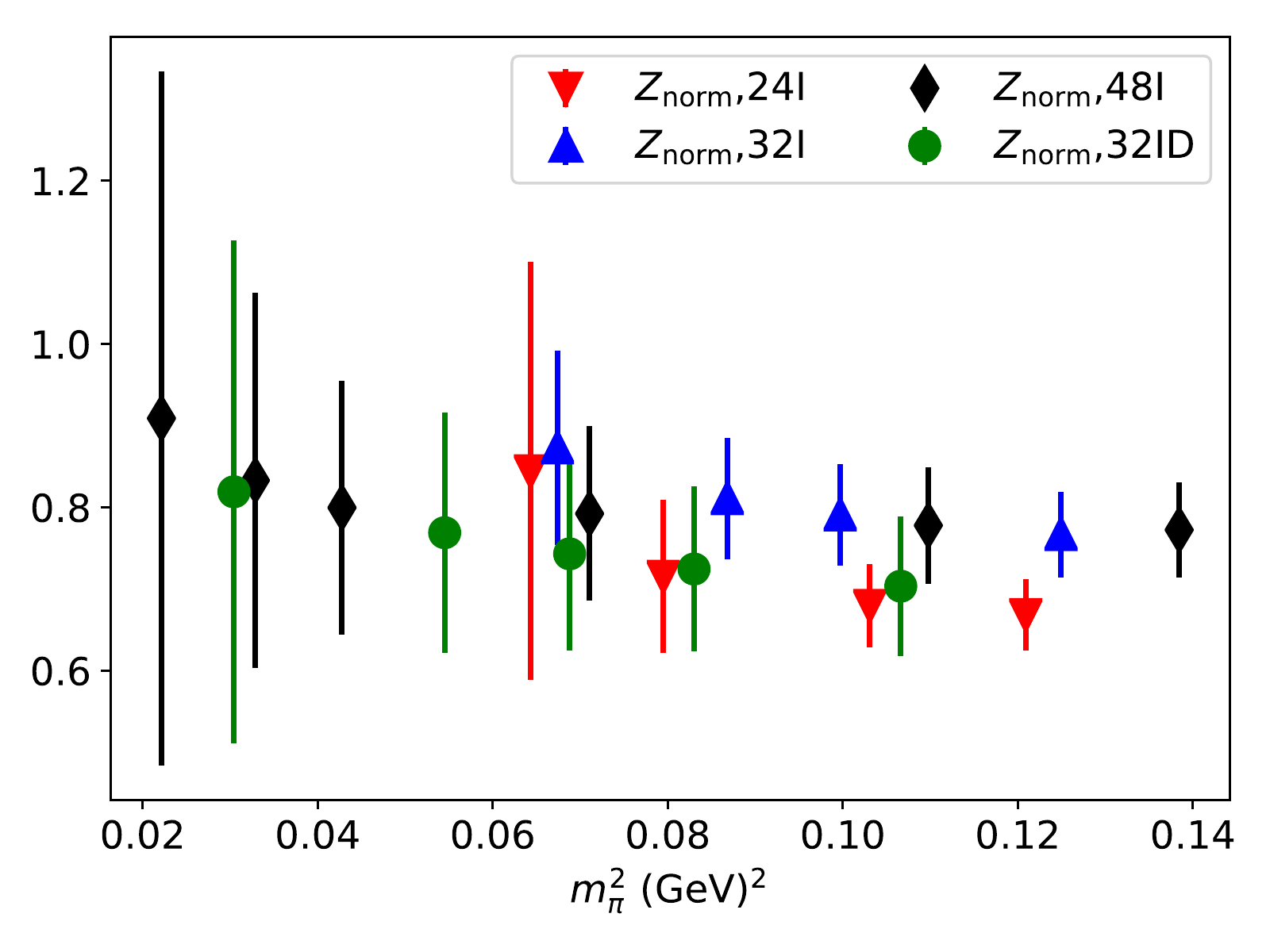}
\caption{The momentum fractions of different quark flavors and glue in the proton, at $\overline{\textrm{MS}}$ 2~GeV. The two upper panels show the $u$, $d$, $s$ and $g$ momentum fractions respectively, and two lower ones show the $u$ - $d$ case (left panel), and also the  normalization factor for the momentum fraction sum rule (right panel). The bands on the figures show our predictions in the continuum limit of the momentum fractions with their statistical (blue) and total (cyan) uncertainties. The data points correspond to our simulation results at different valence quark masses on different ensembles, with the partially quenching effect subtracted.
}
\label{fig:renorm_x}
\end{figure*}

After the renormalization, the total momentum fraction is generally larger than 1 by 20\%--30\% on the four ensembles. Thus the normalization effect here due to the discretization errors is large and we apply a uniform normalization on both the quark and glue momentum fractions at each quark mass of each ensemble, and plot these normalization factors $\bar{Z}=\langle x \rangle_{u+d+s+g}^{-1}$ in the lower-right panel of Fig.~\ref{fig:renorm_x}. 

Then the pion mass dependence of the renormalized and normalized $\langle x \rangle^R_{u,d,s,g}$ are fitted with the following empirical form simultaneously,
\bea
\langle x \rangle^R(m_{\pi}^v,m_{\pi}^{sea},a,L)&=&\langle x \rangle^R_0+D_1((m_{\pi}^v)^2-(m_{\pi}^0)^2)\nonumber\\
&&+D_2\big((m_{\pi}^v)^2-(m_{\pi}^{sea})^2\big)\nonumber\\
&&+D^{I/ID}_{3} a^2+D_4e^{-m_{\pi}^vL},
\eea
and the $\chi^2$/d.o.f. is 0.20. Our prediction of the $\langle x \rangle^R_{u,d,s,g}$ are 0.307(30)(18), 0.160(27)(40), 0.051(26)(5), and 0.482(69)(48) respectively, where the first error is the statistical one and the second error includes the systematic uncertainties from the chiral, continuum, and infinite volume interpolation/extrapolations. The systematic uncertainties from the two-state fit and CDER for $\langle x \rangle_g$ haven't been taken into account yet and will be investigated in the future. With the normalization factor shown in lower-right panel of Fig.~\ref{fig:renorm_x}. All the predictions of the momentum fractions are consistent with the phenomenological global fit at $\overline{\textrm{MS}}$ 2 GeV, e.g., CT14~\cite{Dulat:2015mca} values $\langle x \rangle^R_{u}=0.348(5)$, $\langle x \rangle^R_{d}=0.190(5)$, $\langle x \rangle^R_{s}=0.035(9)$ and  $\langle x \rangle^R_{g}=0.416(9)$. The other global fits results~\cite{Harland-Lang:2014zoa,Abramowicz:2015mha,Accardi:2016qay,Alekhin:2017kpj,Ball:2017nwa} summarized in Ref.~\cite{Lin:2017snn} are consistent with CT14. After the partially quenching effect term proportional to $D_2$ is subtracted, the $\langle x \rangle^R_{u,d,s,g}$ at different ensembles and valence quark masses are illustrated in Fig.~\ref{fig:renorm_x} as a function of $m_{\pi}^2$,  in the upper-left panel for the $u$ and $d$ cases and the upper-right panel for the $s$ and $g$ cases. The bands on the figures show our predictions in the continuum limit with their uncertainties (blue for the statistics and cyan  for the total). 

\begin{figure}[hbp]
\begin{center}
    \includegraphics[scale=0.5]{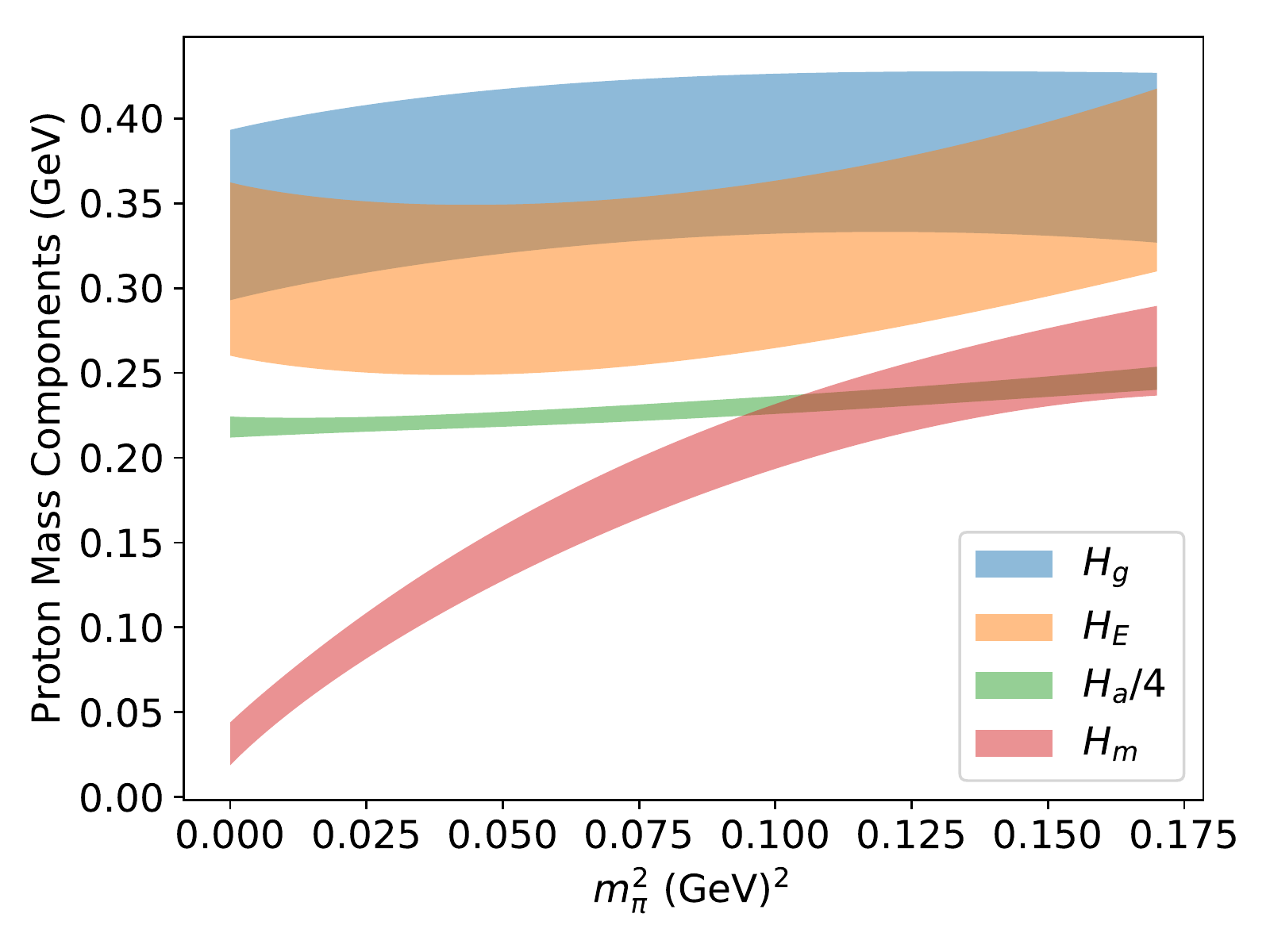}
\end{center}
\caption{The valence pion mass dependence of the proton mass decomposition in terms of the quark condensate ($\langle H_m \rangle$), quark energy $\langle H_E \rangle$, glue field energy $\langle H_g \rangle$ and trace anomaly $\langle H_a \rangle/4$.}
\label{fig:2}
\end{figure}

  We also predict the iso-vector momentum fraction $\langle x \rangle^R_{u-d}$ as 0.151(28)(29), which is consistent with the CT14 result 0.158(6)~\cite{Dulat:2015mca}, in the lower-left panel of Fig.~\ref{fig:renorm_x}.

{\it Final proton mass decomposition:}  With these momentum fractions at $\overline{\textrm{MS}}$ 2 GeV, we can apply Eqs.~(\ref{eq:H_E}) and (\ref{eq:H_g}) to obtain the quark and glue energy contributions in the proton mass (or more precisely, the proton energy in  the rest frame). Combined with the quark scalar condensate and trace anomaly contributions, the entire proton mass decomposition is illustrated in Fig.~\ref{fig:2} as a function of the valence pion mass. As shown in the figure, the major quark mass dependence comes from the quark condensate term, and the other components are almost independent of the quark mass. At the physical point, the quark and glue energy contributions are 32(4)(4)\% and 36(5)(4)\% respectively. With the quark scalar condensate contribution of 9(2)(1)\%~\cite{Yang:2015uis}, we can obtain that a quarter of the trace anomaly contributes 23(1)(1)\% with $N_f=2+1$.\\ 

In summary, we present a simulation strategy to calculate the proton mass decomposition. The renormalization and mixing between the quark 
and glue energy can be calculated non-perturbatively, and the quark scalar condensate contribution and the trace anomaly are renormalization group invariant.  Based on this strategy, the lattice simulation is carried out on four ensembles with three lattice spacings and volumes, and several pion masses including the physical pion mass, to control the respective systematic uncertainties. 
With non-perturbative renormalization and normalization, 
the individual $u,d,s$ and glue momentum fractions agree with those from the global fit in the
$\overline{\textrm{MS}}$ scheme at 2 GeV. Quark energy, gluon energy, and quantum anomaly contributions to the proton mass are fairly insensitive to the pion mass up to 400 MeV within our statistical and systematic uncertainties.

\begin{widetext}
\end{widetext}

\section*{Acknowledgments}

We thank the RBC and UKQCD collaborations for providing
us their DWF gauge configurations. YY is
supported by the US National Science Foundation under
grant PHY 1653405 ``CAREER: Constraining Parton
Distribution Functions for New-Physics Searches." Y.C. and Z.L. acknowledge the support of the National Science Foundation of China under Grants No. 11575196, No. 11575197, No. 11335001.
This work is partially supported by DOE grant DE-SC0013065 and by the DOE TMD topical collaboration. This research
used resources of the Oak Ridge Leadership Computing
Facility at the Oak Ridge National Laboratory,
which is supported by the Office of Science of the U.S.
Department of Energy under Contract No.$~$DE-AC05-00OR22725. This work used Stampede time under the
Extreme Science and Engineering Discovery Environment
(XSEDE), which is supported by National Science Foundation
grant number ACI-1053575. We also thank the National
Energy Research Scientific Computing Center (NERSC)
for providing HPC resources that have contributed to the
research results reported within this paper. We acknowledge
the facilities of the USQCD Collaboration used for
this research in part, which are funded by the Office of
Science of the U.S. Department of Energy.

\begin{widetext}

\section*{Supplementary materials}

\subsection{The non-perturbative renormalization of the quark and gauge EMT}
The renormalized momentum fractions $\langle x \rangle^R$ in the $\overline{\textrm{MS}}$ scheme at scale $\mu$ are
\bea
\langle x \rangle^R_{u,d,s}=Z^{\overline{\textrm{MS}}}_{QQ}(\mu)\langle x \rangle_{u,d,s}+\delta Z^{\overline{\textrm{MS}}}_{QQ}(\mu)\sum_{q=u,d,s} \langle x \rangle_{q}+Z^{\overline{\textrm{MS}}}_{QG}(\mu)\langle x \rangle_g,\ \langle x \rangle^R_g=Z^{\overline{\textrm{MS}}}_{GQ}(\mu)\sum_{q=u,d,s} \langle x \rangle_{q}+Z^{\overline{\textrm{MS}}}_{GG}\langle x \rangle_g,
\eea
where $\langle x \rangle_{u,d,s,g}$ is the bare momentum fraction under the lattice regularization, and 
the renormalization constants at $\overline{\textrm{MS}}$ scale $\mu$ are defined through the RI/MOM scheme at scale $\mu_R$,
\bea
&&\left(\begin{array}{cc}
Z^{\overline{\textrm{MS}}}_{QQ}(\mu)+N_f\delta Z^{\overline{\textrm{MS}}}_{QQ} (\mu)&
N_fZ^{\overline{\textrm{MS}}}_{QG}(\mu)   \\
Z^{\overline{\textrm{MS}}}_{GQ}(\mu) &
Z^{\overline{\textrm{MS}}}_{GG}(\mu)
\end{array}\right)\equiv\left\{\left[\left(\begin{array}{cc}
Z_{QQ}(\mu_R)+N_f\delta Z_{QQ}&
N_fZ_{QG}(\mu_R)\\
Z_{GQ}(\mu_R)&
Z_{GG}(\mu_R)
\end{array}\right)\right.\right.\nonumber\\
&&\quad\quad\quad\quad\quad\quad\quad\quad\quad\quad\quad\quad\quad\quad \left.\left.
\left(\begin{array}{cc}
R_{QQ}(\frac{\mu}{\mu_{R}})+{\cal O}(N_f\alpha_s^2) &
N_fR_{QG}(\frac{\mu}{\mu_{R}})\\
R_{GQ}(\frac{\mu}{\mu_{R}}) &
R_{GG}(\frac{\mu}{\mu_{R}})
\end{array}\right)
\right]|_{a^2\mu_R^2\rightarrow0}\right\}^{-1}\\
&&=\left\{\left(\begin{array}{cc}
\left(Z_{QQ}R_{QQ}\right)+N_f\left(\delta Z_{QQ}R_{QQ}+Z_{QG}R_{GQ}\right)&
N_f\left((Z_{QQ}+N_f\delta R_{QQ}) R_{QG}+Z_{QG}R_{GG}\right)\\
\left(Z_{GQ}R_{QQ}+Z_{GG}R_{GQ}\right) &
\left(N_fZ_{GQ} R_{QG}+Z_{GG}R_{GG}\right)
\end{array}\right)(\mu_R,\frac{\mu}{\mu_R})|_{a^2\mu_R^2\rightarrow0}\right\}^{-1}
\label{eq:Z_extrapolation}
\eea
and $Z_{QQ}(\mu)=\left[\left(Z_{QQ}R_{QQ}\right)(\mu_R,\frac{\mu}{\mu_R})|_{a^2\mu_R^2\rightarrow 0}\right]^{-1}$. 

In the above equations, the RI/MOM renormalization constants $Z$ of $\overline{T}^{q,g}_{44}$ are defined with the following conditions suggested by Ref.~\cite{Gracey:2003mr} for cases with quark external legs:
\bea
Z_{QQ}(\mu_R)&=&\frac{V\textrm{Tr}\left[\Gamma^{q}_{\mu\mu}\bar{S}_q^{-1}(p)\left\langle \sum_w \gamma_5 S_q^{\dagger}(p,w) \gamma_5 \gamma_{\mu}\overleftrightarrow{D}_{\mu} S_q(p,w) \right\rangle \bar{S}_q^{-1}(p)\right]}{\left[-i\Gamma^{q}_{\mu\mu}(\gamma_{\mu}\tilde{p}_{\mu}-\frac{1}{4}\pslash{\tilde{p}})\right]Z_q}|_{p^2=\mu_R^2},\label{eq:conditions_qq}\\
\delta Z_{QQ}(\mu_R)&=&\frac{V\textrm{Tr}\left[\Gamma^{q}_{\mu\mu}\bar{S}_q^{-1}(p)\left\langle \overline{T}^q_{\mu\mu} S_q(p) \right\rangle \bar{S}_q^{-1}(p)\right]}{\left[-i\Gamma^{q}_{\mu\mu}(\gamma_{\mu}\tilde{p}_{\mu}-\frac{1}{4}\pslash{\tilde{p}})\right]Z_q}|_{p^2=\mu_R^2},\label{eq:conditions_qq_di}\\ 
Z_{GQ}(\mu_R)&=&\frac{V\textrm{Tr}\left[\Gamma^{q}_{\mu\mu}\bar{S}_q^{-1}(p)\left\langle \overline{T}^g_{\mu\mu} S_q(p) \right\rangle \bar{S}_q^{-1}(p)\right]}{\left[-i\Gamma^{q}_{\mu\mu}(\gamma_{\mu}\tilde{p}_{\mu}-\frac{1}{4}\pslash{\tilde{p}})\right]Z_q}|_{p^2=\mu_R^2},\label{eq:conditions_Gq}
\eea
where $V$ is the lattice volume, $p$ is the momentum of the external quark/gluon state, $\tilde{p}_{\mu}=\textrm{sin}p_{\mu}$, $\Gamma^q_{\mu\mu}=i\gamma_{\mu}\tilde{p}_{\mu}-i\frac{\tilde{p}^2_{\mu}}{\tilde{p}^2}\pslash{\tilde{p}}$ as suggested by Ref.~\cite{Alexandrou:2010me},  the quark propagators are $\bar{S}_q(p)\equiv \langle S_q(p)\rangle \equiv \langle \sum_{x} e^{ipx}S_q(p,x)\rangle$ with $S_q(p,x)=\sum_{y} e^{-ipy}\psi(x)\bar{\psi}(y) $,  and $Z_q$ is defined through the axial-vector vertex correction and Ward identity~\cite{Liu:2013yxz}. Note that the index $\mu$ in Eqs.~(\ref{eq:conditions_qq})-(\ref{eq:conditions_Gq}) is not summed while the results with different values of $\mu$ can be averaged.  

For the case of gluon external legs, the definitions are the following, as inspired by Refs.~\cite{Yang:2016xsb,Yang:2018bft},
\bea
Z_{QG}(\mu_R)&=&\xi^{-1}Z_{b}(\mu_R, \overline{T}^q)-(\xi^{-1}-1)Z_{a}(\mu_R, \overline{T}^q),\label{eq:conditions_qG}\\ 
Z_{GG}(\mu_R)&=&\xi^{-1}Z_{b}(\mu_R, \overline{T}^g)-(\xi^{-1}-1)Z_{a}(\mu_R, \overline{T}^g),\label{eq:conditions_GG}
\eea
where $\xi\equiv\frac{\sum_{\mu}p_{\mu}^4}{(\sum_{\mu}p_{\mu}^2)^2}$, and $Z_{a/b}$ are defined by
\bea\label{eq:glue_prj_new}
Z_{a}(\mu_R, \overline{T})&=&\frac{p^2\langle (k_{\mu}\overline{T}_{\mu\nu}q_{\nu})\Tr[A_{\rho}(p) A_{\tau}(-p)\Gamma_{\rho\tau}]\rangle}
{2k^2q^2\langle\Tr[A_{\rho}(p) A_{\tau}(-p)\Gamma_{\rho\tau}]\rangle}|_{\tiny{\substack{p^2=\mu_R^2,  k+q=p, k\cdot q=0, \\ \Gamma_{\rho\tau}=\delta_{\rho\tau}-\frac{k_{\rho}k_{\tau}}{k^2}-\frac{q_{\rho}q_{\tau}}{q^2} }}}\\
Z_{b}(\mu_R, \overline{T})&=&\frac{\langle (p_{\mu}\overline{T}_{\mu\nu}p_{\nu}-l_{\mu}\overline{T}_{\mu\nu}l_{\nu})\Tr[A_{\rho}(p) A_{\tau}(-p)\tilde{\Gamma}_{\rho\tau}]\rangle}
{2p^2\langle\Tr[A_{\rho}(p) A_{\tau}(-p)\tilde{\Gamma}_{\rho\tau}]\rangle}|_{\tiny{\substack{p^2=\mu_R^2,  l^2=p^2, l\cdot p=0, \\ \tilde{\Gamma}_{\rho\tau}=\delta_{\rho\tau}-\frac{p_{\rho}p_{\tau}}{p^2}-\frac{l_{\rho}l_{\tau}}{l^2} }}},
\eea
with all the repeated indices summed.

The 3-loop level result of the matching coefficient $R_{QQ}(\frac{\mu}{\mu_{R}})=1+\frac{g^2}{16\pi^2}C_F[\frac{8}{3}\textrm{log}(\mu^2/\mu_R^2)+\frac{31}{9}]+{\cal O}(\alpha_s^2)$ has been obtained in Ref.~\cite{Gracey:2003mr}, while just the 1-loop level results of the other $R$'s are available~\cite{Yang:2016xsb}:
\bea
R_{QG}&=&-\frac{g^2}{16\pi^2}[\frac{2}{3}\textrm{log}(\mu^2/\mu_R^2)+\frac{4}{9}]+{\cal O}(\alpha_s^2),\ R_{GQ}=-\frac{g^2C_F}{16\pi^2}[\frac{8}{3}\textrm{log}(\mu^2/\mu_R^2)+\frac{22}{9}]+{\cal O}(\alpha_s^2), \nonumber\\
R_{GG}&=&1+\frac{g^2N_f}{16\pi^2}[\frac{2}{3}\textrm{log}(\mu^2/\mu_R^2)+\frac{10}{9}]-\frac{g^2N_c}{16\pi^2}\frac{5}{12}+{\cal O}(\alpha_s^2).
\eea
Thus we will use the 3-loop matching for the renormalization of the quark momentum fraction and keep all the matching at 1-loop level in the rest of the renormalization calculation.

In the practical lattice calculation, all the $Z$'s suffer from discretization errors and we need to repeat the calculation of $Z$ at different $p^2$, match them to the $\overline{\textrm{MS}}$ scheme at the $\mu_R$ scale, evolve them from $\mu_R$ to a fixed scale such as \mbox{2~GeV}, and then apply the $a^2\mu_R^2=a^2p^2$ extrapolation to get the final result of $Z$. To apply this extrapolation properly, we should calculate the vertex corrections in the quark and gluon states with exactly the same momenta, and combine them first. But it is not necessary for most of the cases except for the quark to gluon mixing, as we will discuss case by case.

 \begin{figure*}[htb!] 
\includegraphics[scale=0.7]{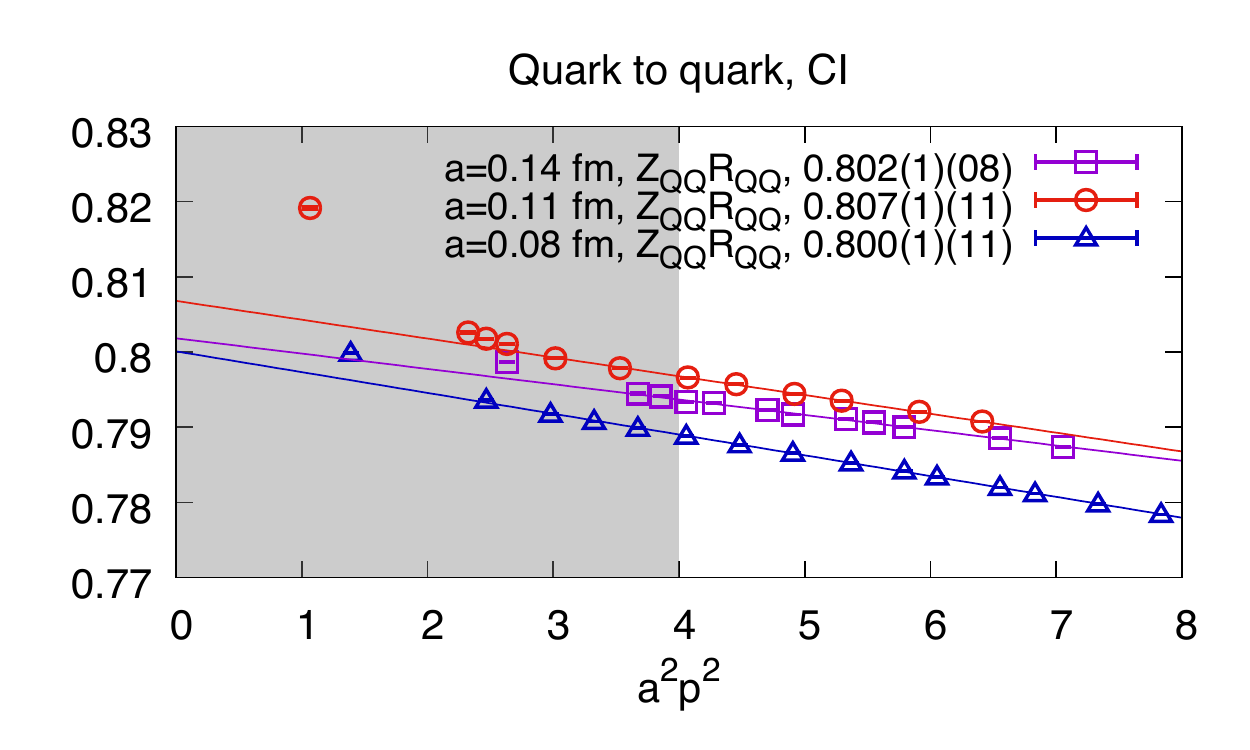}
\includegraphics[scale=0.7]{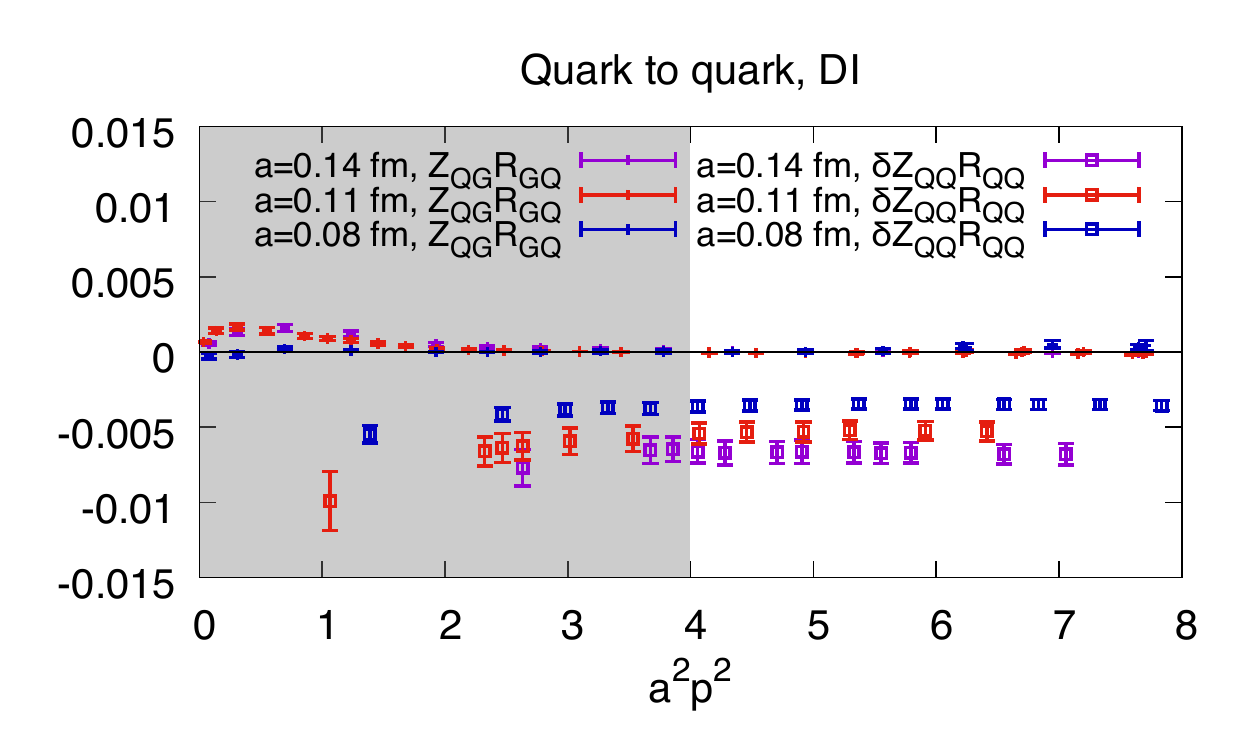}\\
\caption{The quark EMT renormalization constant $Z_{QQ}R_{QQ}$ (left panel) and the disconnected pieces ($\delta Z_{QQ}R_{QQ}$ and $Z_{QG}R_{GQ}$, in the right panel) at different lattice spacings at $\overline{\textrm{MS}}$ scheme 2 GeV, as a function of $a^2p^2$. With the momenta along the body-diagonal direction, the discretization errors are small and the results have a mild $a^2p^2$ dependence.}
\label{fig:Z_q}
\end{figure*}

\subsection{Quark EMT renormalization}

\begin{table}[htbp]
\begin{center}
\caption{\label{table:Z_q} Uncertainties (in percent) of $Z^{\overline{\textrm{MS}}}_{QQ}$ in the chiral limit, at different lattice spacings.}
\begin{tabular}{c|ccc|ccc}
operator & \multicolumn{3}{c|}{$\overline{T}^{q}_{\mu\mu}$}&  \multicolumn{3}{c}{$\overline{T}^{q}_{\mu\nu}$} \\
\hline
a (fm) & 0.143 & 0.111 & 0.083 & 0.143 & 0.111 & 0.083 \\
\hline
        Statistical uncertainty& 0.12 & 0.07 & 0.10 & 0.28 & 0.15 & 0.20\\
        \hline
        Conversional ratio($Z_{\overline{T}^{q}_{\mu\nu}}^{\overline{\text{MS}}}/Z_{\overline{T}^{q}_{\mu\nu}}^{\text{RI'}}$) & 0.87 & 0.87 & 0.87 & 0.87 & 0.87 & 0.87\\
        Conversional ratio($Z_{q}^{\text{RI'}}/Z_q^{\text{RI}}$) & 0.12 & 0.12 & 0.12 & 0.12 & 0.12 & 0.12 \\
        $\Lambda_{\text{QCD}}^{\overline{\textrm{MS}}}$ & 0.53 & 0.26 & 0.02 & 0.53 & 0.26 & 0.03\\
        Perturbative running & $<$ 0.01 &  $<$ 0.01 &  $<$ 0.01 &  $<$ 0.01 &  $<$ 0.01 &  $<$ 0.01\\
        Lattice spacing & 0.16 & 0.08 & 0.08 & 0.16 & 0.08 & 0.08\\  
        Fit range of $a^2p^2$ & 0.08 & 0.02 & 0.01 & 0.24 & 0.37 & 0.11\\
        $m_s^{sea}\neq 0$ & $\sim$0.25 & $\sim$1 & $\sim$1 & $\sim$0.25 & $\sim$1 & $\sim$1  \\
        \hline
        Total systematic uncertainty & $\sim 1.07$ & $\sim 1.36$ & $\sim 1.33$ &$\sim 1.09$ & $\sim 1.41$ & $\sim 1.34$ \\\end{tabular}
\end{center}
\end{table}

For the quark external state, we choose 12 momenta with $\frac{\sum_{\mu}p_{\mu}^4}{(\sum_{\mu}p_{\mu}^2)^2}<0.27$ for 30 configurations and 6 valence quark masses at each lattice spacing, and use Landau gauge fixed momentum volume sources to improve the signal. With the given momenta, we extrapolated the result to the chiral limit and then applied the continuum matching to get the value at $\overline{\textrm{MS}}$  2~GeV. The values of $Z_{QQ}R_{QQ}$ for three lattice spacings are plotted in the left panel of Fig.~\ref{fig:Z_q}, and the values at $a^2p^2=0$ limit are obtained based on linear extrapolation of the data in the range $a^2p^2\in[4,8]$.  It turns out that the $a^2p^2$ dependence is small by using $\tilde{p}_{\mu}$ in the definition of the tree level operator and projection, when the chosen momenta are along the body-diagonal direction. The calculation on the 48I ensemble is skipped as its lattice spacing is almost the same as that of the 24I ensemble. 

We followed the same strategy used in Ref.~\cite{Liu:2013yxz,Bi:2017ybi} to analyze the systematic uncertainties and the error budgets are collected in Tab.~\ref{table:Z_q}. The systematic uncertainty of $m_s^{sea}\neq 0$ are estimated as $\sim$1\% on the 24I and 32I ensembles and $\sim$0.25\% on 32ID ensemble, based on our previous study with multiple sea quark masses~\cite{Liu:2013yxz}.  The values for the off-diagonal parts of the quark EMT cases at $a$=0.143, 0.111 and 0.083 fm (at $a^2p^2=0$ limit) are slightly different. They are 0.786(2)(9), 0.796(1)(11) and 0.793(1)(11) respectively.  

In the right panel of Fig.~\ref{fig:Z_q}, the other two terms needed by the singlet quark EMT renormalization, $\delta Z_{QQ}R_{QQ}$ and $Z_{QG}R_{GQ}$ for the disconnected contribution, are plotted. Regardless of the simulation details of the $Z_{QG}$ which will be addressed later, the contribution of second term is much smaller than the first term, and thus the $a^2p^2$ extrapolation can be carried out with the $\delta Z_{QQ}R_{QQ}$ term alone in the $a^2p^2\in[4,8]$ range, considering the value of the other term as a systematic uncertainty which is $\sim$0.001. Note that the CDER technique \cite{Liu:2017man} is applied to the correlation function of $\delta Z_{QQ}$ with the cutoff $r_0\sim$ 1 fm,
\bea\label{eq:quark_cder}
\left\langle \textrm{Tr}[\overline{T}^{q}_{\mu\mu}] S(p) \right\rangle=\left\langle \int \textrm{d}^4x \textrm{d}^{4}y \textrm{Tr}[\overline{T}^{q,g}_{\mu\mu}](x) S(p,y) \right\rangle\simeq \left\langle \int \textrm{d}^4x \int_{r\le r_0}\textrm{d}^{4}r \textrm{Tr}[\overline{T}^{q}_{\mu\mu}](x) S(p,x+r) \right\rangle
\eea

\subsection{Gauge EMT renormalization}

In our previous investigation of the glue EMT renormalization~\cite{Yang:2018bft}, we chose the momenta with two transverse components to calculate  $Z_{GG}$:
\bea
Z_{GG}(\mu_R)&=&\frac{p^2\langle (\overline{T}^{g}_{\mu\mu}-\overline{T}^{g}_{\nu\nu})\Tr[A_{\rho}(p) A_{\rho}(-p)]\rangle}
{2p^2_{\mu}\langle\Tr[A_{\rho}(p) A_{\rho}(-p)]\rangle}|_{\tiny{\substack{p^2=\mu_R^2,\\ \rho\neq\mu\neq\nu, \\p_\rho=0,\\p_\nu=0}}},\label{eq:conditions_GG1}
\eea
But the discretization errors at the range $a^2p^2\in[4,8]$ used in the quark external legs are too large to carry out a reliable fit, when we want to combine it with $Z_{GQ}$. Thus in this work, we introduce two improvements to suppress the $a^2p^2$ corrections:

1. The $a^2p^2$ correction from the HYP smearing on the glue operator: We found that most of the $a^2p^2$ correction in $Z_{GG}$ with the HYP smeared gauge EMT can be removed by the following ratio $f(a^2p^2)$,
\bea\label{eq:Zg_improvement1}
\tilde{Z}_{GG}(a^2p^2)=Z_{GG}(a^2p^2)f(a^2p^2\rightarrow 0)/f(a^2p^2),\ f(a^2p^2)=\frac{\langle\Tr[A^{HYP}_{\rho}(p) A^{HYP}_{\rho}(-p)]\rangle}{\langle\Tr[A_{\rho}(p) A_{\rho}(-p)]\rangle},
\eea
where $A_{\rho}^{HYP}=\sum_{x} e^{ip\cdot(x+\frac{1}{2}\hat{\rho})}\left[\frac{{\cal U}_\rho(x)-{\cal U}^{\dagger}_\rho(x)}{2ig_0a}\right]_\text{traceless}$ is the HYP-smeared gauge potential defined from the HYP-smeared gauge link ${\cal U}_\rho$, and $f(a^2p^2)$ is the ratio of two propagators $S(p)\equiv\langle\Tr[A_{\rho}(p) A_{\rho}(-p)]\rangle$ with and without HYP smearing. Since the scale dependence of $f(a^2p^2)$ is cancelled, only the $a^2p^2$ discretization errors exist up to the normalization $f(a^2p^2\rightarrow 0)$. Note that such a normalization corresponds to the tadpole effect of the HYP smearing which should be included in the renormalization constant $Z_{GG}$. Since the value of $S(p)$ doesn't exist at $p^2=0$, we will have to fit $f(a^2p^2)$ with a polynomial of $a^2p^2$ to get the value of $f(a^2p^2\rightarrow 0)$.

2. We also extend the calculation of the $Z_{GG}$ to the momenta along the body-diagonal direction, by considering the projected renormalization constants defined in Eq.~\ref{eq:conditions_GG}. 
They can be rewritten into the combination of the renormalization constant of the traceless diagonal gauge EMT $Z_{GG}$, and the off-diagonal one $Z_{GG}^{off}$ by
\bea\label{eq:Zg_improvement2}
Z_{a}(\mu_R, \overline{T}_g)=Z^{off}_{GG}(\mu_R), Z_{b}(\mu_R, \overline{T}_g)=Z^{off}_{GG}(\mu_R)+\xi(Z_{GG}-Z^{off}_{GG})(\mu_R).
\eea

 \begin{figure*}[htb!] 
 \includegraphics[scale=0.7]{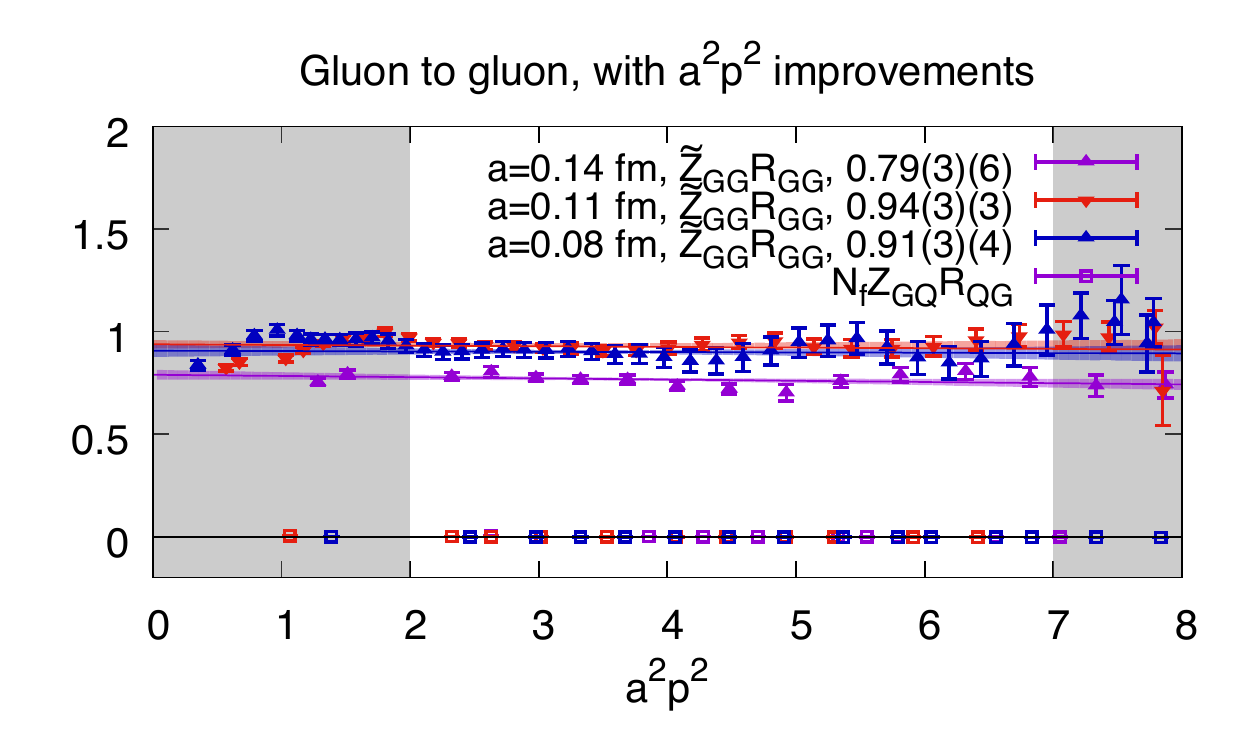}
\includegraphics[scale=0.7]{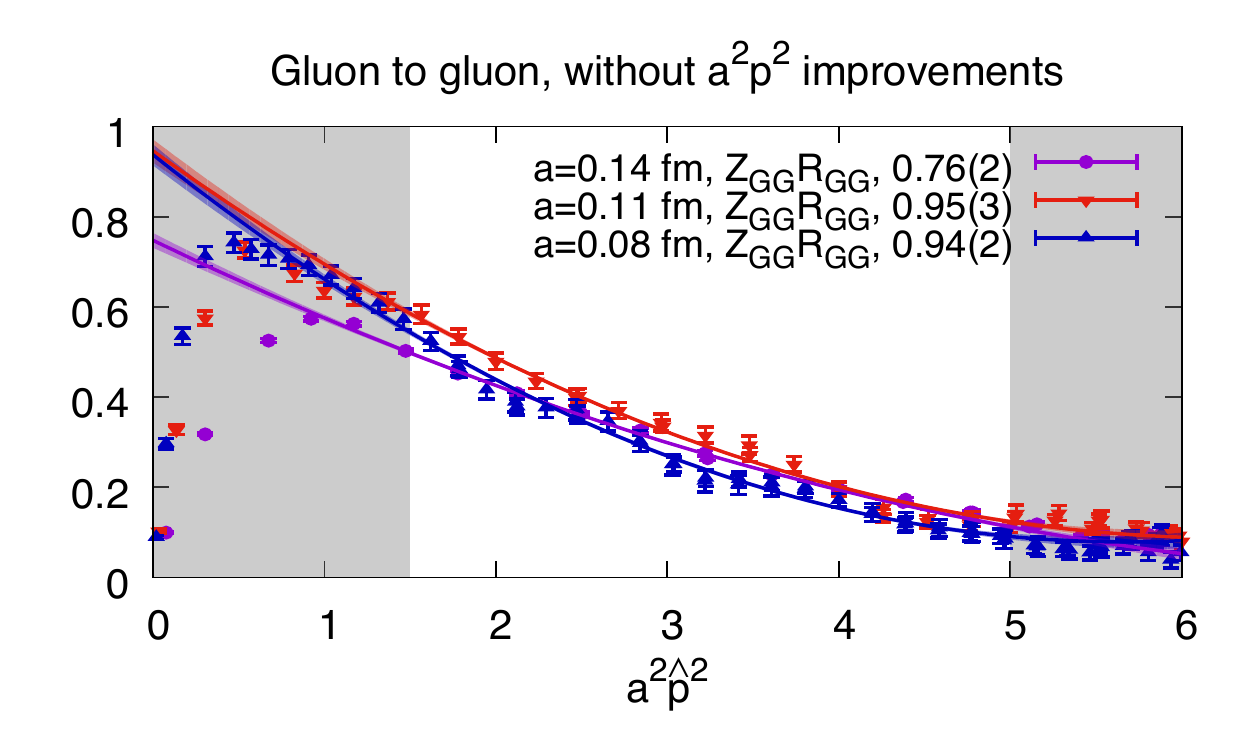}
\caption{The gauge EMT renormalization constant, with (left panel) and without (right panel) the $a^2p^2$ improvements described in Eq.~(\ref{eq:Zg_improvement1}) and (\ref{eq:Zg_improvement2}). The contribution from $\tilde{Z}^{RI}_{GG}R_{GG}$ is dominant and that from  $N_fZ_{GQ} R_{QG}$ is negligible. The $a^2p^2\rightarrow0$ results with and without improvement can provide consistent predictions.}
\label{fig:Z_g}
\end{figure*}

With 949/356/290 configurations respectively, the values of $\tilde{Z}^{RI}_{GG}R_{GG}$ on 32ID/48I/64I ensembles with 1 step of HYP smearing are illustrated in the left panel of Fig.~\ref{fig:Z_g}
. Note that the ensemble 64I, which has almost the same lattice spacing as 32I (0.0837(2) {\it vs.} 0.0828(3)) but with a larger volume and physical pion mass, is used for the calculation, as a similar calculation on 32I would require $\sim$5,000 configurations, which are not available, to reach the similar accuracy. After both improvements above are applied, the $a^2p^2$ dependences of the $\tilde{Z}^{RI}_{GG}R_{GG}$ are mild. The contributions from the other term $N_fZ_{GQ} R_{QG}$ are also illustrated on the same figures. Both the values and their uncertainties ($\sim$0.005) are much smaller than the statistical uncertainty of $\tilde{Z}^{RI}_{GG}$ and thus can be dropped safely. With linear extrapolation of the data in the range 
$a^2p^2\in[2,7]$ and the polynomial fit of the $f(a^2p^2)$ in the same range, we obtain the renormalization constants at $a$=0.143, 0.114 and 0.084 fm to be 0.79(3)(6), 0.94(3)(3), and 0.91(3)(4) respectively, with the second error determined from varying the starting/ending points of the $a^2p^2$ range by 1. 

For comparison, we also illustrate the results obtained from the previous strategy used in Ref.~\cite{Yang:2018bft} in the right panel of Fig.~\ref{fig:Z_g}, with the fit of the data in the range of $a^2\hat{p}^2\equiv(\textrm{sin}\frac{pa}{2})^2\in[1.5, 5]$. The $a^2p^2\rightarrow0$ results of based on the polynomial fit are consistent with the present linear-fit ones. 

\subsection{Mixings}

 \begin{figure*}[htb!] 
\includegraphics[scale=0.70]{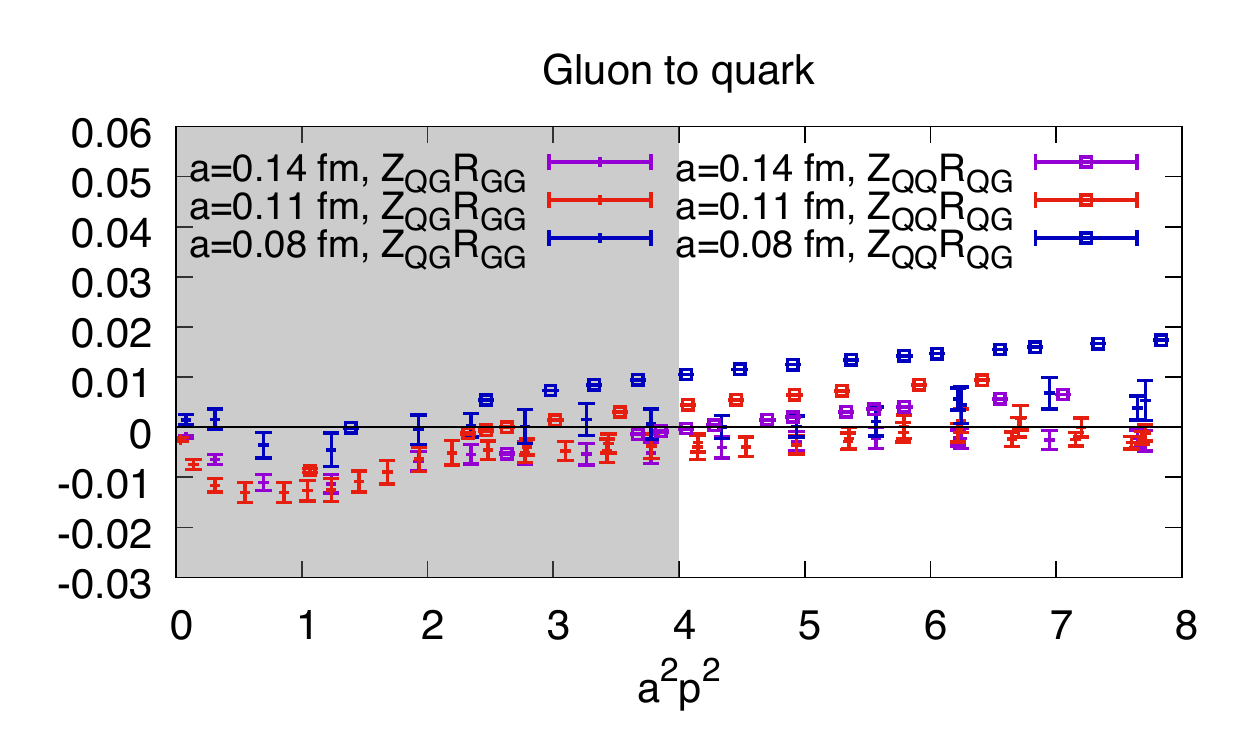}
\includegraphics[scale=0.70]{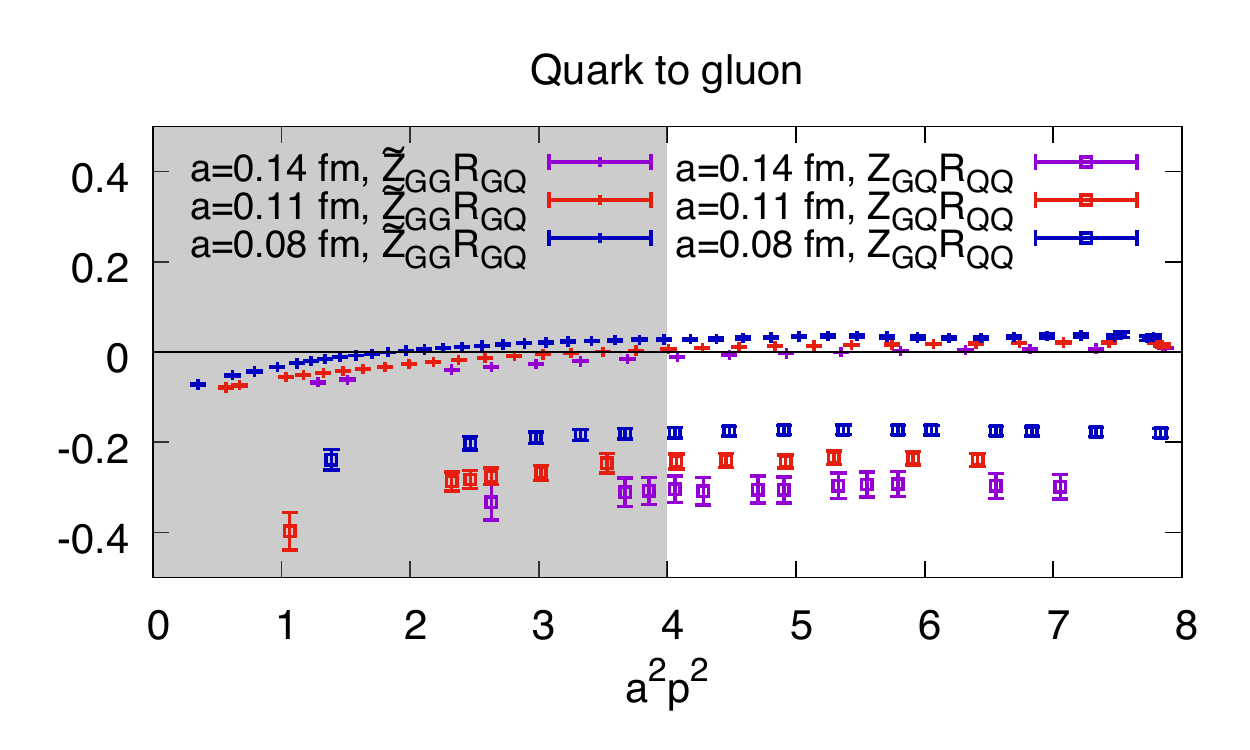}
\caption{The renormalization matrix element on three ensembles with different lattice spacings, as a function of $a^2p^2$.}
\label{fig:Z_mix}
\end{figure*}

For the mixing from glue to quark, both the contributions from $Z_{QQ} R_{QG}$ and $Z_{QG}R_{GG}$ are small, as illustrated in the left panel of Fig.~\ref{fig:Z_mix}. The total contribution can be estimated as $-$0.010(10),  $-$0.005(10), $-$0.000(10) per flavor at $a$=0.143, 0.111 and 0.083 fm respectively, when the linear extrapolation is applied in the range $a^2p^2\in[4,8]$. 

The last piece of the jigsaw is the mixing from quark to glue. The CDER technique used for $\delta Z_{QQ}$ as in Eq.~\ref{eq:quark_cder} can be applied for the calculation here with the gauge EMT operator.  As in the right panel of Fig.~\ref{fig:Z_mix}, the contributions from $\tilde{Z}^{RI}_{GG}R_{GQ}$ are just a few percent, while those from $Z_{GQ}R_{QQ}$ are sizable. Since the $a^2p^2$ dependence of $\tilde{Z}^{RI}_{GG}$ is small as in the left panel of Fig.~\ref{fig:Z_g}, we can approximate the $\tilde{Z}^{RI}_{GG}(a^2p^2)$ by its $a^2p^2$ extrapolated value and assign $\sim$ 3\% systematic uncertainty from the difference between the values at $a^2p^2$ equal to 4 and 8. After combining the correction and applying the linear extrapolation in the range $a^2p^2\in[4,8]$, we obtain the total mixing at $a$=0.143, 0.111 and 0.083 fm as $-$0.34(2)(4), $-$0.26(2)(3), and $-$0.13(2)(1) respectively. 

\end{widetext}

\bibliography{reference.bib}

\begin{thebibliography}{32}%
\makeatletter
\providecommand \@ifxundefined [1]{%
 \@ifx{#1\undefined}
}%
\providecommand \@ifnum [1]{%
 \ifnum #1\expandafter \@firstoftwo
 \else \expandafter \@secondoftwo
 \fi
}%
\providecommand \@ifx [1]{%
 \ifx #1\expandafter \@firstoftwo
 \else \expandafter \@secondoftwo
 \fi
}%
\providecommand \natexlab [1]{#1}%
\providecommand \enquote  [1]{``#1''}%
\providecommand \bibnamefont  [1]{#1}%
\providecommand \bibfnamefont [1]{#1}%
\providecommand \citenamefont [1]{#1}%
\providecommand \href@noop [0]{\@secondoftwo}%
\providecommand \href [0]{\begingroup \@sanitize@url \@href}%
\providecommand \@href[1]{\@@startlink{#1}\@@href}%
\providecommand \@@href[1]{\endgroup#1\@@endlink}%
\providecommand \@sanitize@url [0]{\catcode `\\12\catcode `\$12\catcode
  `\&12\catcode `\#12\catcode `\^12\catcode `\_12\catcode `\%12\relax}%
\providecommand \@@startlink[1]{}%
\providecommand \@@endlink[0]{}%
\providecommand \url  [0]{\begingroup\@sanitize@url \@url }%
\providecommand \@url [1]{\endgroup\@href {#1}{\urlprefix }}%
\providecommand \urlprefix  [0]{URL }%
\providecommand \Eprint [0]{\href }%
\providecommand \doibase [0]{http://dx.doi.org/}%
\providecommand \selectlanguage [0]{\@gobble}%
\providecommand \bibinfo  [0]{\@secondoftwo}%
\providecommand \bibfield  [0]{\@secondoftwo}%
\providecommand \translation [1]{[#1]}%
\providecommand \BibitemOpen [0]{}%
\providecommand \bibitemStop [0]{}%
\providecommand \bibitemNoStop [0]{.\EOS\space}%
\providecommand \EOS [0]{\spacefactor3000\relax}%
\providecommand \BibitemShut  [1]{\csname bibitem#1\endcsname}%
\let\auto@bib@innerbib\@empty
%</preamble>
\bibitem [{\citenamefont {Ji}(1995)}]{Ji:1994av}%
  \BibitemOpen
  \bibfield  {author} {\bibinfo {author} {\bibfnamefont {X.-D.}\ \bibnamefont
  {Ji}},\ }\href {\doibase 10.1103/PhysRevLett.74.1071} {\bibfield  {journal}
  {\bibinfo  {journal} {Phys. Rev. Lett.}\ }\textbf {\bibinfo {volume} {74}},\
  \bibinfo {pages} {1071} (\bibinfo {year} {1995})},\ \Eprint
  {http://arxiv.org/abs/hep-ph/9410274} {arXiv:hep-ph/9410274 [hep-ph]}
  \BibitemShut {NoStop}%
%%CITATION = HEP-PH/9410274;%%
\bibitem [{\citenamefont {Yang}\ \emph {et~al.}(2015)\citenamefont {Yang},
  \citenamefont {Chen}, \citenamefont {Draper}, \citenamefont {Gong},
  \citenamefont {Liu}, \citenamefont {Liu},\ and\ \citenamefont
  {Ma}}]{Yang:2014xsa}%
  \BibitemOpen
  \bibfield  {author} {\bibinfo {author} {\bibfnamefont {Y.-B.}\ \bibnamefont
  {Yang}}, \bibinfo {author} {\bibfnamefont {Y.}~\bibnamefont {Chen}}, \bibinfo
  {author} {\bibfnamefont {T.}~\bibnamefont {Draper}}, \bibinfo {author}
  {\bibfnamefont {M.}~\bibnamefont {Gong}}, \bibinfo {author} {\bibfnamefont
  {K.-F.}\ \bibnamefont {Liu}}, \bibinfo {author} {\bibfnamefont
  {Z.}~\bibnamefont {Liu}}, \ and\ \bibinfo {author} {\bibfnamefont {J.-P.}\
  \bibnamefont {Ma}},\ }\href {\doibase 10.1103/PhysRevD.91.074516} {\bibfield
  {journal} {\bibinfo  {journal} {Phys. Rev.}\ }\textbf {\bibinfo {volume}
  {D91}},\ \bibinfo {pages} {074516} (\bibinfo {year} {2015})},\ \Eprint
  {http://arxiv.org/abs/1405.4440} {arXiv:1405.4440 [hep-ph]} \BibitemShut
  {NoStop}%
%%CITATION = ARXIV:1405.4440;%%
\bibitem [{\citenamefont {Yang}\ \emph
  {et~al.}(2016{\natexlab{a}})\citenamefont {Yang}, \citenamefont {Alexandru},
  \citenamefont {Draper}, \citenamefont {Liang},\ and\ \citenamefont
  {Liu}}]{Yang:2015uis}%
  \BibitemOpen
  \bibfield  {author} {\bibinfo {author} {\bibfnamefont {Y.-B.}\ \bibnamefont
  {Yang}}, \bibinfo {author} {\bibfnamefont {A.}~\bibnamefont {Alexandru}},
  \bibinfo {author} {\bibfnamefont {T.}~\bibnamefont {Draper}}, \bibinfo
  {author} {\bibfnamefont {J.}~\bibnamefont {Liang}}, \ and\ \bibinfo {author}
  {\bibfnamefont {K.-F.}\ \bibnamefont {Liu}} (\bibinfo {collaboration}
  {xQCD}),\ }\href {\doibase 10.1103/PhysRevD.94.054503} {\bibfield  {journal}
  {\bibinfo  {journal} {Phys. Rev.}\ }\textbf {\bibinfo {volume} {D94}},\
  \bibinfo {pages} {054503} (\bibinfo {year} {2016}{\natexlab{a}})},\ \Eprint
  {http://arxiv.org/abs/1511.09089} {arXiv:1511.09089 [hep-lat]} \BibitemShut
  {NoStop}%
%%CITATION = ARXIV:1511.09089;%%
\bibitem [{\citenamefont {Aoki}\ \emph {et~al.}(2011)\citenamefont {Aoki} \emph
  {et~al.}}]{Aoki:2010dy}%
  \BibitemOpen
  \bibfield  {author} {\bibinfo {author} {\bibfnamefont {Y.}~\bibnamefont
  {Aoki}} \emph {et~al.} (\bibinfo {collaboration} {RBC, UKQCD}),\ }\href
  {\doibase 10.1103/PhysRevD.83.074508} {\bibfield  {journal} {\bibinfo
  {journal} {Phys. Rev.}\ }\textbf {\bibinfo {volume} {D83}},\ \bibinfo {pages}
  {074508} (\bibinfo {year} {2011})},\ \Eprint {http://arxiv.org/abs/1011.0892}
  {arXiv:1011.0892 [hep-lat]} \BibitemShut {NoStop}%
%%CITATION = ARXIV:1011.0892;%%
\bibitem [{\citenamefont {Blum}\ \emph {et~al.}(2016)\citenamefont {Blum} \emph
  {et~al.}}]{Blum:2014tka}%
  \BibitemOpen
  \bibfield  {author} {\bibinfo {author} {\bibfnamefont {T.}~\bibnamefont
  {Blum}} \emph {et~al.} (\bibinfo {collaboration} {RBC, UKQCD}),\ }\href
  {\doibase 10.1103/PhysRevD.93.074505} {\bibfield  {journal} {\bibinfo
  {journal} {Phys. Rev.}\ }\textbf {\bibinfo {volume} {D93}},\ \bibinfo {pages}
  {074505} (\bibinfo {year} {2016})},\ \Eprint {http://arxiv.org/abs/1411.7017}
  {arXiv:1411.7017 [hep-lat]} \BibitemShut {NoStop}%
%%CITATION = ARXIV:1411.7017;%%
\bibitem [{\citenamefont {Sufian}\ \emph {et~al.}(2017)\citenamefont {Sufian},
  \citenamefont {Yang}, \citenamefont {Alexandru}, \citenamefont {Draper},
  \citenamefont {Liang},\ and\ \citenamefont {Liu}}]{Sufian:2016pex}%
  \BibitemOpen
  \bibfield  {author} {\bibinfo {author} {\bibfnamefont {R.~S.}\ \bibnamefont
  {Sufian}}, \bibinfo {author} {\bibfnamefont {Y.-B.}\ \bibnamefont {Yang}},
  \bibinfo {author} {\bibfnamefont {A.}~\bibnamefont {Alexandru}}, \bibinfo
  {author} {\bibfnamefont {T.}~\bibnamefont {Draper}}, \bibinfo {author}
  {\bibfnamefont {J.}~\bibnamefont {Liang}}, \ and\ \bibinfo {author}
  {\bibfnamefont {K.-F.}\ \bibnamefont {Liu}},\ }\href {\doibase
  10.1103/PhysRevLett.118.042001} {\bibfield  {journal} {\bibinfo  {journal}
  {Phys. Rev. Lett.}\ }\textbf {\bibinfo {volume} {118}},\ \bibinfo {pages}
  {042001} (\bibinfo {year} {2017})},\ \Eprint
  {http://arxiv.org/abs/1606.07075} {arXiv:1606.07075 [hep-ph]} \BibitemShut
  {NoStop}%
%%CITATION = ARXIV:1606.07075;%%
\bibitem [{\citenamefont {Chiu}(1999)}]{Chiu:1998eu}%
  \BibitemOpen
  \bibfield  {author} {\bibinfo {author} {\bibfnamefont {T.-W.}\ \bibnamefont
  {Chiu}},\ }\href {\doibase 10.1103/PhysRevD.60.034503} {\bibfield  {journal}
  {\bibinfo  {journal} {Phys. Rev.}\ }\textbf {\bibinfo {volume} {D60}},\
  \bibinfo {pages} {034503} (\bibinfo {year} {1999})},\ \Eprint
  {http://arxiv.org/abs/hep-lat/9810052} {arXiv:hep-lat/9810052 [hep-lat]}
  \BibitemShut {NoStop}%
%%CITATION = HEP-LAT/9810052;%%
\bibitem [{\citenamefont {Liu}(2005)}]{Liu:2002qu}%
  \BibitemOpen
  \bibfield  {author} {\bibinfo {author} {\bibfnamefont {K.-F.}\ \bibnamefont
  {Liu}},\ }\href {\doibase 10.1142/S0217751X05022366} {\bibfield  {journal}
  {\bibinfo  {journal} {Int. J. Mod. Phys.}\ }\textbf {\bibinfo {volume}
  {A20}},\ \bibinfo {pages} {7241} (\bibinfo {year} {2005})},\ \Eprint
  {http://arxiv.org/abs/hep-lat/0206002} {arXiv:hep-lat/0206002 [hep-lat]}
  \BibitemShut {NoStop}%
%%CITATION = HEP-LAT/0206002;%%
\bibitem [{\citenamefont {Chiu}\ and\ \citenamefont
  {Zenkin}(1999)}]{Chiu:1998gp}%
  \BibitemOpen
  \bibfield  {author} {\bibinfo {author} {\bibfnamefont {T.-W.}\ \bibnamefont
  {Chiu}}\ and\ \bibinfo {author} {\bibfnamefont {S.~V.}\ \bibnamefont
  {Zenkin}},\ }\href {\doibase 10.1103/PhysRevD.59.074501} {\bibfield
  {journal} {\bibinfo  {journal} {Phys. Rev.}\ }\textbf {\bibinfo {volume}
  {D59}},\ \bibinfo {pages} {074501} (\bibinfo {year} {1999})},\ \Eprint
  {http://arxiv.org/abs/hep-lat/9806019} {arXiv:hep-lat/9806019 [hep-lat]}
  \BibitemShut {NoStop}%
%%CITATION = HEP-LAT/9806019;%%
\bibitem [{\citenamefont {Li}\ \emph {et~al.}(2010)\citenamefont {Li} \emph
  {et~al.}}]{Li:2010pw}%
  \BibitemOpen
  \bibfield  {author} {\bibinfo {author} {\bibfnamefont {A.}~\bibnamefont {Li}}
  \emph {et~al.} (\bibinfo {collaboration} {$\chi$QCD}),\ }\href {\doibase
  10.1103/PhysRevD.82.114501} {\bibfield  {journal} {\bibinfo  {journal} {Phys.
  Rev.}\ }\textbf {\bibinfo {volume} {D82}},\ \bibinfo {pages} {114501}
  (\bibinfo {year} {2010})},\ \Eprint {http://arxiv.org/abs/1005.5424}
  {arXiv:1005.5424 [hep-lat]} \BibitemShut {NoStop}%
%%CITATION = ARXIV:1005.5424;%%
\bibitem [{\citenamefont {Gong}\ \emph {et~al.}(2013)\citenamefont {Gong} \emph
  {et~al.}}]{Gong:2013vja}%
  \BibitemOpen
  \bibfield  {author} {\bibinfo {author} {\bibfnamefont {M.}~\bibnamefont
  {Gong}} \emph {et~al.} (\bibinfo {collaboration} {$\chi$QCD}),\ }\href
  {\doibase 10.1103/PhysRevD.88.014503} {\bibfield  {journal} {\bibinfo
  {journal} {Phys. Rev.}\ }\textbf {\bibinfo {volume} {D88}},\ \bibinfo {pages}
  {014503} (\bibinfo {year} {2013})},\ \Eprint {http://arxiv.org/abs/1304.1194}
  {arXiv:1304.1194 [hep-ph]} \BibitemShut {NoStop}%
%%CITATION = ARXIV:1304.1194;%%
\bibitem [{\citenamefont {Yang}\ \emph
  {et~al.}(2016{\natexlab{b}})\citenamefont {Yang}, \citenamefont {Alexandru},
  \citenamefont {Draper}, \citenamefont {Gong},\ and\ \citenamefont
  {Liu}}]{Yang:2015zja}%
  \BibitemOpen
  \bibfield  {author} {\bibinfo {author} {\bibfnamefont {Y.-B.}\ \bibnamefont
  {Yang}}, \bibinfo {author} {\bibfnamefont {A.}~\bibnamefont {Alexandru}},
  \bibinfo {author} {\bibfnamefont {T.}~\bibnamefont {Draper}}, \bibinfo
  {author} {\bibfnamefont {M.}~\bibnamefont {Gong}}, \ and\ \bibinfo {author}
  {\bibfnamefont {K.-F.}\ \bibnamefont {Liu}},\ }\href {\doibase
  10.1103/PhysRevD.93.034503} {\bibfield  {journal} {\bibinfo  {journal} {Phys.
  Rev.}\ }\textbf {\bibinfo {volume} {D93}},\ \bibinfo {pages} {034503}
  (\bibinfo {year} {2016}{\natexlab{b}})},\ \Eprint
  {http://arxiv.org/abs/1509.04616} {arXiv:1509.04616 [hep-lat]} \BibitemShut
  {NoStop}%
%%CITATION = ARXIV:1509.04616;%%
\bibitem [{\citenamefont {Liang}\ \emph {et~al.}(2017)\citenamefont {Liang},
  \citenamefont {Yang}, \citenamefont {Liu}, \citenamefont {Alexandru},
  \citenamefont {Draper},\ and\ \citenamefont {Sufian}}]{Liang:2016fgy}%
  \BibitemOpen
  \bibfield  {author} {\bibinfo {author} {\bibfnamefont {J.}~\bibnamefont
  {Liang}}, \bibinfo {author} {\bibfnamefont {Y.-B.}\ \bibnamefont {Yang}},
  \bibinfo {author} {\bibfnamefont {K.-F.}\ \bibnamefont {Liu}}, \bibinfo
  {author} {\bibfnamefont {A.}~\bibnamefont {Alexandru}}, \bibinfo {author}
  {\bibfnamefont {T.}~\bibnamefont {Draper}}, \ and\ \bibinfo {author}
  {\bibfnamefont {R.~S.}\ \bibnamefont {Sufian}},\ }\href {\doibase
  10.1103/PhysRevD.96.034519} {\bibfield  {journal} {\bibinfo  {journal} {Phys.
  Rev.}\ }\textbf {\bibinfo {volume} {D96}},\ \bibinfo {pages} {034519}
  (\bibinfo {year} {2017})},\ \Eprint {http://arxiv.org/abs/1612.04388}
  {arXiv:1612.04388 [hep-lat]} \BibitemShut {NoStop}%
%%CITATION = ARXIV:1612.04388;%%
\bibitem [{\citenamefont {Tiburzi}(2005)}]{Tiburzi:2005is}%
  \BibitemOpen
  \bibfield  {author} {\bibinfo {author} {\bibfnamefont {B.~C.}\ \bibnamefont
  {Tiburzi}},\ }\href {\doibase 10.1103/PhysRevD.72.094501,
  10.1103/PhysRevD.79.039904} {\bibfield  {journal} {\bibinfo  {journal} {Phys.
  Rev.}\ }\textbf {\bibinfo {volume} {D72}},\ \bibinfo {pages} {094501}
  (\bibinfo {year} {2005})},\ \bibinfo {note} {[Erratum: Phys.
  Rev.D79,039904(2009)]},\ \Eprint {http://arxiv.org/abs/hep-lat/0508019}
  {arXiv:hep-lat/0508019 [hep-lat]} \BibitemShut {NoStop}%
%%CITATION = HEP-LAT/0508019;%%
\bibitem [{\citenamefont {Patrignani}\ \emph {et~al.}(2016)\citenamefont
  {Patrignani} \emph {et~al.}}]{Patrignani:2016xqp}%
  \BibitemOpen
  \bibfield  {author} {\bibinfo {author} {\bibfnamefont {C.}~\bibnamefont
  {Patrignani}} \emph {et~al.} (\bibinfo {collaboration} {Particle Data
  Group}),\ }\href {\doibase 10.1088/1674-1137/40/10/100001} {\bibfield
  {journal} {\bibinfo  {journal} {Chin. Phys.}\ }\textbf {\bibinfo {volume}
  {C40}},\ \bibinfo {pages} {100001} (\bibinfo {year} {2016})}\BibitemShut
  {NoStop}%
%%CITATION = CHPHD,C40,100001;%%
\bibitem [{\citenamefont {Horsley}\ \emph {et~al.}(2012)\citenamefont
  {Horsley}, \citenamefont {Millo}, \citenamefont {Nakamura}, \citenamefont
  {Perlt}, \citenamefont {Pleiter}, \citenamefont {Rakow}, \citenamefont
  {Schierholz}, \citenamefont {Schiller}, \citenamefont {Winter},\ and\
  \citenamefont {Zanotti}}]{Horsley:2012pz}%
  \BibitemOpen
  \bibfield  {author} {\bibinfo {author} {\bibfnamefont {R.}~\bibnamefont
  {Horsley}}, \bibinfo {author} {\bibfnamefont {R.}~\bibnamefont {Millo}},
  \bibinfo {author} {\bibfnamefont {Y.}~\bibnamefont {Nakamura}}, \bibinfo
  {author} {\bibfnamefont {H.}~\bibnamefont {Perlt}}, \bibinfo {author}
  {\bibfnamefont {D.}~\bibnamefont {Pleiter}}, \bibinfo {author} {\bibfnamefont
  {P.~E.~L.}\ \bibnamefont {Rakow}}, \bibinfo {author} {\bibfnamefont
  {G.}~\bibnamefont {Schierholz}}, \bibinfo {author} {\bibfnamefont
  {A.}~\bibnamefont {Schiller}}, \bibinfo {author} {\bibfnamefont
  {F.}~\bibnamefont {Winter}}, \ and\ \bibinfo {author} {\bibfnamefont {J.~M.}\
  \bibnamefont {Zanotti}} (\bibinfo {collaboration} {UKQCD, QCDSF}),\ }\href
  {\doibase 10.1016/j.physletb.2012.07.004} {\bibfield  {journal} {\bibinfo
  {journal} {Phys. Lett.}\ }\textbf {\bibinfo {volume} {B714}},\ \bibinfo
  {pages} {312} (\bibinfo {year} {2012})},\ \Eprint
  {http://arxiv.org/abs/1205.6410} {arXiv:1205.6410 [hep-lat]} \BibitemShut
  {NoStop}%
%%CITATION = ARXIV:1205.6410;%%
\bibitem [{\citenamefont {Yang}\ \emph {et~al.}(2017)\citenamefont {Yang},
  \citenamefont {Sufian}, \citenamefont {Alexandru}, \citenamefont {Draper},
  \citenamefont {Glatzmaier}, \citenamefont {Liu},\ and\ \citenamefont
  {Zhao}}]{Yang:2016plb}%
  \BibitemOpen
  \bibfield  {author} {\bibinfo {author} {\bibfnamefont {Y.-B.}\ \bibnamefont
  {Yang}}, \bibinfo {author} {\bibfnamefont {R.~S.}\ \bibnamefont {Sufian}},
  \bibinfo {author} {\bibfnamefont {A.}~\bibnamefont {Alexandru}}, \bibinfo
  {author} {\bibfnamefont {T.}~\bibnamefont {Draper}}, \bibinfo {author}
  {\bibfnamefont {M.~J.}\ \bibnamefont {Glatzmaier}}, \bibinfo {author}
  {\bibfnamefont {K.-F.}\ \bibnamefont {Liu}}, \ and\ \bibinfo {author}
  {\bibfnamefont {Y.}~\bibnamefont {Zhao}},\ }\href {\doibase
  10.1103/PhysRevLett.118.102001} {\bibfield  {journal} {\bibinfo  {journal}
  {Phys. Rev. Lett.}\ }\textbf {\bibinfo {volume} {118}},\ \bibinfo {pages}
  {102001} (\bibinfo {year} {2017})},\ \Eprint
  {http://arxiv.org/abs/1609.05937} {arXiv:1609.05937 [hep-ph]} \BibitemShut
  {NoStop}%
%%CITATION = ARXIV:1609.05937;%%
\bibitem [{\citenamefont {Liu}\ \emph {et~al.}(2018)\citenamefont {Liu},
  \citenamefont {Liang},\ and\ \citenamefont {Yang}}]{Liu:2017man}%
  \BibitemOpen
  \bibfield  {author} {\bibinfo {author} {\bibfnamefont {K.-F.}\ \bibnamefont
  {Liu}}, \bibinfo {author} {\bibfnamefont {J.}~\bibnamefont {Liang}}, \ and\
  \bibinfo {author} {\bibfnamefont {Y.-B.}\ \bibnamefont {Yang}},\ }\href
  {\doibase 10.1103/PhysRevD.97.034507} {\bibfield  {journal} {\bibinfo
  {journal} {Phys. Rev.}\ }\textbf {\bibinfo {volume} {D97}},\ \bibinfo {pages}
  {034507} (\bibinfo {year} {2018})},\ \Eprint
  {http://arxiv.org/abs/1705.06358} {arXiv:1705.06358 [hep-lat]} \BibitemShut
  {NoStop}%
%%CITATION = ARXIV:1705.06358;%%
\bibitem [{\citenamefont {Yang}\ \emph {et~al.}(2018)\citenamefont {Yang},
  \citenamefont {Gong}, \citenamefont {Liang}, \citenamefont {Lin},
  \citenamefont {Liu}, \citenamefont {Pefkou},\ and\ \citenamefont
  {Shanahan}}]{Yang:2018bft}%
  \BibitemOpen
  \bibfield  {author} {\bibinfo {author} {\bibfnamefont {Y.-B.}\ \bibnamefont
  {Yang}}, \bibinfo {author} {\bibfnamefont {M.}~\bibnamefont {Gong}}, \bibinfo
  {author} {\bibfnamefont {J.}~\bibnamefont {Liang}}, \bibinfo {author}
  {\bibfnamefont {H.-W.}\ \bibnamefont {Lin}}, \bibinfo {author} {\bibfnamefont
  {K.-F.}\ \bibnamefont {Liu}}, \bibinfo {author} {\bibfnamefont
  {D.}~\bibnamefont {Pefkou}}, \ and\ \bibinfo {author} {\bibfnamefont
  {P.}~\bibnamefont {Shanahan}},\ }\href@noop {} {\  (\bibinfo {year}
  {2018})},\ \Eprint {http://arxiv.org/abs/1805.00531} {arXiv:1805.00531
  [hep-lat]} \BibitemShut {NoStop}%
%%CITATION = ARXIV:1805.00531;%%
\bibitem [{\citenamefont {Gracey}(2003)}]{Gracey:2003mr}%
  \BibitemOpen
  \bibfield  {author} {\bibinfo {author} {\bibfnamefont {J.~A.}\ \bibnamefont
  {Gracey}},\ }\href {\doibase 10.1016/S0550-3213(03)00543-1} {\bibfield
  {journal} {\bibinfo  {journal} {Nucl. Phys.}\ }\textbf {\bibinfo {volume}
  {B667}},\ \bibinfo {pages} {242} (\bibinfo {year} {2003})},\ \Eprint
  {http://arxiv.org/abs/hep-ph/0306163} {arXiv:hep-ph/0306163 [hep-ph]}
  \BibitemShut {NoStop}%
%%CITATION = HEP-PH/0306163;%%
\bibitem [{\citenamefont {Yang}\ \emph
  {et~al.}(2016{\natexlab{c}})\citenamefont {Yang}, \citenamefont {Glatzmaier},
  \citenamefont {Liu},\ and\ \citenamefont {Zhao}}]{Yang:2016xsb}%
  \BibitemOpen
  \bibfield  {author} {\bibinfo {author} {\bibfnamefont {Y.-B.}\ \bibnamefont
  {Yang}}, \bibinfo {author} {\bibfnamefont {M.}~\bibnamefont {Glatzmaier}},
  \bibinfo {author} {\bibfnamefont {K.-F.}\ \bibnamefont {Liu}}, \ and\
  \bibinfo {author} {\bibfnamefont {Y.}~\bibnamefont {Zhao}},\ }\href@noop {}
  {\  (\bibinfo {year} {2016}{\natexlab{c}})},\ \Eprint
  {http://arxiv.org/abs/1612.02855} {arXiv:1612.02855 [nucl-th]} \BibitemShut
  {NoStop}%
%%CITATION = ARXIV:1612.02855;%%
\bibitem [{yan()}]{yang:2018NPR}%
  \BibitemOpen
  \href@noop {} {\bibinfo  {journal} {Supplementary materials}\ }\BibitemShut
  {NoStop}%
\bibitem [{\citenamefont {Dulat}\ \emph {et~al.}(2016)\citenamefont {Dulat},
  \citenamefont {Hou}, \citenamefont {Gao}, \citenamefont {Guzzi},
  \citenamefont {Huston}, \citenamefont {Nadolsky}, \citenamefont {Pumplin},
  \citenamefont {Schmidt}, \citenamefont {Stump},\ and\ \citenamefont
  {Yuan}}]{Dulat:2015mca}%
  \BibitemOpen
\bibfield  {journal} {  }\bibfield  {author} {\bibinfo {author} {\bibfnamefont
  {S.}~\bibnamefont {Dulat}}, \bibinfo {author} {\bibfnamefont {T.-J.}\
  \bibnamefont {Hou}}, \bibinfo {author} {\bibfnamefont {J.}~\bibnamefont
  {Gao}}, \bibinfo {author} {\bibfnamefont {M.}~\bibnamefont {Guzzi}}, \bibinfo
  {author} {\bibfnamefont {J.}~\bibnamefont {Huston}}, \bibinfo {author}
  {\bibfnamefont {P.}~\bibnamefont {Nadolsky}}, \bibinfo {author}
  {\bibfnamefont {J.}~\bibnamefont {Pumplin}}, \bibinfo {author} {\bibfnamefont
  {C.}~\bibnamefont {Schmidt}}, \bibinfo {author} {\bibfnamefont
  {D.}~\bibnamefont {Stump}}, \ and\ \bibinfo {author} {\bibfnamefont {C.~P.}\
  \bibnamefont {Yuan}},\ }\href {\doibase 10.1103/PhysRevD.93.033006}
  {\bibfield  {journal} {\bibinfo  {journal} {Phys. Rev.}\ }\textbf {\bibinfo
  {volume} {D93}},\ \bibinfo {pages} {033006} (\bibinfo {year} {2016})},\
  \Eprint {http://arxiv.org/abs/1506.07443} {arXiv:1506.07443 [hep-ph]}
  \BibitemShut {NoStop}%
%%CITATION = ARXIV:1506.07443;%%
\bibitem [{\citenamefont {Harland-Lang}\ \emph {et~al.}(2015)\citenamefont
  {Harland-Lang}, \citenamefont {Martin}, \citenamefont {Motylinski},\ and\
  \citenamefont {Thorne}}]{Harland-Lang:2014zoa}%
  \BibitemOpen
  \bibfield  {author} {\bibinfo {author} {\bibfnamefont {L.~A.}\ \bibnamefont
  {Harland-Lang}}, \bibinfo {author} {\bibfnamefont {A.~D.}\ \bibnamefont
  {Martin}}, \bibinfo {author} {\bibfnamefont {P.}~\bibnamefont {Motylinski}},
  \ and\ \bibinfo {author} {\bibfnamefont {R.~S.}\ \bibnamefont {Thorne}},\
  }\href {\doibase 10.1140/epjc/s10052-015-3397-6} {\bibfield  {journal}
  {\bibinfo  {journal} {Eur. Phys. J.}\ }\textbf {\bibinfo {volume} {C75}},\
  \bibinfo {pages} {204} (\bibinfo {year} {2015})},\ \Eprint
  {http://arxiv.org/abs/1412.3989} {arXiv:1412.3989 [hep-ph]} \BibitemShut
  {NoStop}%
%%CITATION = ARXIV:1412.3989;%%
\bibitem [{\citenamefont {Abramowicz}\ \emph {et~al.}(2015)\citenamefont
  {Abramowicz} \emph {et~al.}}]{Abramowicz:2015mha}%
  \BibitemOpen
  \bibfield  {author} {\bibinfo {author} {\bibfnamefont {H.}~\bibnamefont
  {Abramowicz}} \emph {et~al.} (\bibinfo {collaboration} {ZEUS, H1}),\ }\href
  {\doibase 10.1140/epjc/s10052-015-3710-4} {\bibfield  {journal} {\bibinfo
  {journal} {Eur. Phys. J.}\ }\textbf {\bibinfo {volume} {C75}},\ \bibinfo
  {pages} {580} (\bibinfo {year} {2015})},\ \Eprint
  {http://arxiv.org/abs/1506.06042} {arXiv:1506.06042 [hep-ex]} \BibitemShut
  {NoStop}%
%%CITATION = ARXIV:1506.06042;%%
\bibitem [{\citenamefont {Accardi}\ \emph {et~al.}(2016)\citenamefont
  {Accardi}, \citenamefont {Brady}, \citenamefont {Melnitchouk}, \citenamefont
  {Owens},\ and\ \citenamefont {Sato}}]{Accardi:2016qay}%
  \BibitemOpen
  \bibfield  {author} {\bibinfo {author} {\bibfnamefont {A.}~\bibnamefont
  {Accardi}}, \bibinfo {author} {\bibfnamefont {L.~T.}\ \bibnamefont {Brady}},
  \bibinfo {author} {\bibfnamefont {W.}~\bibnamefont {Melnitchouk}}, \bibinfo
  {author} {\bibfnamefont {J.~F.}\ \bibnamefont {Owens}}, \ and\ \bibinfo
  {author} {\bibfnamefont {N.}~\bibnamefont {Sato}},\ }\href {\doibase
  10.1103/PhysRevD.93.114017} {\bibfield  {journal} {\bibinfo  {journal} {Phys.
  Rev.}\ }\textbf {\bibinfo {volume} {D93}},\ \bibinfo {pages} {114017}
  (\bibinfo {year} {2016})},\ \Eprint {http://arxiv.org/abs/1602.03154}
  {arXiv:1602.03154 [hep-ph]} \BibitemShut {NoStop}%
%%CITATION = ARXIV:1602.03154;%%
\bibitem [{\citenamefont {Alekhin}\ \emph {et~al.}(2017)\citenamefont
  {Alekhin}, \citenamefont {Bl{\"u}mlein}, \citenamefont {Moch},\ and\
  \citenamefont {Placakyte}}]{Alekhin:2017kpj}%
  \BibitemOpen
  \bibfield  {author} {\bibinfo {author} {\bibfnamefont {S.}~\bibnamefont
  {Alekhin}}, \bibinfo {author} {\bibfnamefont {J.}~\bibnamefont
  {Bl{\"u}mlein}}, \bibinfo {author} {\bibfnamefont {S.}~\bibnamefont {Moch}},
  \ and\ \bibinfo {author} {\bibfnamefont {R.}~\bibnamefont {Placakyte}},\
  }\href {\doibase 10.1103/PhysRevD.96.014011} {\bibfield  {journal} {\bibinfo
  {journal} {Phys. Rev.}\ }\textbf {\bibinfo {volume} {D96}},\ \bibinfo {pages}
  {014011} (\bibinfo {year} {2017})},\ \Eprint
  {http://arxiv.org/abs/1701.05838} {arXiv:1701.05838 [hep-ph]} \BibitemShut
  {NoStop}%
%%CITATION = ARXIV:1701.05838;%%
\bibitem [{\citenamefont {Ball}\ \emph {et~al.}(2017)\citenamefont {Ball} \emph
  {et~al.}}]{Ball:2017nwa}%
  \BibitemOpen
  \bibfield  {author} {\bibinfo {author} {\bibfnamefont {R.~D.}\ \bibnamefont
  {Ball}} \emph {et~al.} (\bibinfo {collaboration} {NNPDF}),\ }\href {\doibase
  10.1140/epjc/s10052-017-5199-5} {\bibfield  {journal} {\bibinfo  {journal}
  {Eur. Phys. J.}\ }\textbf {\bibinfo {volume} {C77}},\ \bibinfo {pages} {663}
  (\bibinfo {year} {2017})},\ \Eprint {http://arxiv.org/abs/1706.00428}
  {arXiv:1706.00428 [hep-ph]} \BibitemShut {NoStop}%
%%CITATION = ARXIV:1706.00428;%%
\bibitem [{\citenamefont {Lin}\ \emph {et~al.}(2018)\citenamefont {Lin} \emph
  {et~al.}}]{Lin:2017snn}%
  \BibitemOpen
  \bibfield  {author} {\bibinfo {author} {\bibfnamefont {H.-W.}\ \bibnamefont
  {Lin}} \emph {et~al.},\ }\href {\doibase 10.1016/j.ppnp.2018.01.007}
  {\bibfield  {journal} {\bibinfo  {journal} {Prog. Part. Nucl. Phys.}\
  }\textbf {\bibinfo {volume} {100}},\ \bibinfo {pages} {107} (\bibinfo {year}
  {2018})},\ \Eprint {http://arxiv.org/abs/1711.07916} {arXiv:1711.07916
  [hep-ph]} \BibitemShut {NoStop}%
%%CITATION = ARXIV:1711.07916;%%
\bibitem [{\citenamefont {Alexandrou}\ \emph {et~al.}(2011)\citenamefont
  {Alexandrou}, \citenamefont {Constantinou}, \citenamefont {Korzec},
  \citenamefont {Panagopoulos},\ and\ \citenamefont
  {Stylianou}}]{Alexandrou:2010me}%
  \BibitemOpen
  \bibfield  {author} {\bibinfo {author} {\bibfnamefont {C.}~\bibnamefont
  {Alexandrou}}, \bibinfo {author} {\bibfnamefont {M.}~\bibnamefont
  {Constantinou}}, \bibinfo {author} {\bibfnamefont {T.}~\bibnamefont
  {Korzec}}, \bibinfo {author} {\bibfnamefont {H.}~\bibnamefont
  {Panagopoulos}}, \ and\ \bibinfo {author} {\bibfnamefont {F.}~\bibnamefont
  {Stylianou}},\ }\href {\doibase 10.1103/PhysRevD.83.014503} {\bibfield
  {journal} {\bibinfo  {journal} {Phys. Rev.}\ }\textbf {\bibinfo {volume}
  {D83}},\ \bibinfo {pages} {014503} (\bibinfo {year} {2011})},\ \Eprint
  {http://arxiv.org/abs/1006.1920} {arXiv:1006.1920 [hep-lat]} \BibitemShut
  {NoStop}%
%%CITATION = ARXIV:1006.1920;%%
\bibitem [{\citenamefont {Liu}\ \emph {et~al.}(2014)\citenamefont {Liu},
  \citenamefont {Chen}, \citenamefont {Dong}, \citenamefont {Glatzmaier},
  \citenamefont {Gong}, \citenamefont {Li}, \citenamefont {Liu}, \citenamefont
  {Yang},\ and\ \citenamefont {Zhang}}]{Liu:2013yxz}%
  \BibitemOpen
  \bibfield  {author} {\bibinfo {author} {\bibfnamefont {Z.}~\bibnamefont
  {Liu}}, \bibinfo {author} {\bibfnamefont {Y.}~\bibnamefont {Chen}}, \bibinfo
  {author} {\bibfnamefont {S.-J.}\ \bibnamefont {Dong}}, \bibinfo {author}
  {\bibfnamefont {M.}~\bibnamefont {Glatzmaier}}, \bibinfo {author}
  {\bibfnamefont {M.}~\bibnamefont {Gong}}, \bibinfo {author} {\bibfnamefont
  {A.}~\bibnamefont {Li}}, \bibinfo {author} {\bibfnamefont {K.-F.}\
  \bibnamefont {Liu}}, \bibinfo {author} {\bibfnamefont {Y.-B.}\ \bibnamefont
  {Yang}}, \ and\ \bibinfo {author} {\bibfnamefont {J.-B.}\ \bibnamefont
  {Zhang}} (\bibinfo {collaboration} {chiQCD}),\ }\href {\doibase
  10.1103/PhysRevD.90.034505} {\bibfield  {journal} {\bibinfo  {journal} {Phys.
  Rev.}\ }\textbf {\bibinfo {volume} {D90}},\ \bibinfo {pages} {034505}
  (\bibinfo {year} {2014})},\ \Eprint {http://arxiv.org/abs/1312.7628}
  {arXiv:1312.7628 [hep-lat]} \BibitemShut {NoStop}%
%%CITATION = ARXIV:1312.7628;%%
\bibitem [{\citenamefont {Bi}\ \emph {et~al.}(2018)\citenamefont {Bi},
  \citenamefont {Cai}, \citenamefont {Chen}, \citenamefont {Gong},
  \citenamefont {Liu}, \citenamefont {Liu},\ and\ \citenamefont
  {Yang}}]{Bi:2017ybi}%
  \BibitemOpen
  \bibfield  {author} {\bibinfo {author} {\bibfnamefont {Y.}~\bibnamefont
  {Bi}}, \bibinfo {author} {\bibfnamefont {H.}~\bibnamefont {Cai}}, \bibinfo
  {author} {\bibfnamefont {Y.}~\bibnamefont {Chen}}, \bibinfo {author}
  {\bibfnamefont {M.}~\bibnamefont {Gong}}, \bibinfo {author} {\bibfnamefont
  {K.-F.}\ \bibnamefont {Liu}}, \bibinfo {author} {\bibfnamefont
  {Z.}~\bibnamefont {Liu}}, \ and\ \bibinfo {author} {\bibfnamefont {Y.-B.}\
  \bibnamefont {Yang}},\ }\href {\doibase 10.1103/PhysRevD.97.094501}
  {\bibfield  {journal} {\bibinfo  {journal} {Phys. Rev.}\ }\textbf {\bibinfo
  {volume} {D97}},\ \bibinfo {pages} {094501} (\bibinfo {year} {2018})},\
  \Eprint {http://arxiv.org/abs/1710.08678} {arXiv:1710.08678 [hep-lat]}
  \BibitemShut {NoStop}%
%%CITATION = ARXIV:1710.08678;%%
\end{thebibliography}%

%%%%%%%%%%%%%%%%%%%%%%%%%%%%%%%%%%%%%%%%%%%%%%%%%%%%%%%%%%%%%%%%%%%%%%%%%%%%%
\end{document}